\magnification1200
\def\vts{\mkern1mu}
\def\ts{\mkern2mu}
\def\rd{{\rm d}}
\def\ri{{\rm i}}
\def\id{{\rm id}\vts  }
\def\Spin{{\rm Spin}\vts }
\def\Pin{{\rm Pin}\vts }
\def\SO{{\rm SO}\vts }
\def\Ort{{\rm O}\vts }
\def\GL{{\rm GL}\vts }
\def\vol{{\rm vol}\vts }
\def\Cl{{\rm Cl}\vts }
\def\Ad{{\rm Ad}\vts }
\def\End{{\rm End}\ts}
\def\Hom{{\rm Hom}\vts }
\mathchardef\varPhi="0108
\mathchardef\varPsi="0109
\mathchardef\varLambda="0103
\def\lto{\relbar\joinrel\rightarrow}

\catcode`\@=11
\def\undefine#1{\let#1\undefined}
\def\newsymbol#1#2#3#4#5{\let\next@\relax
 \ifnum#2=\@ne\let\next@\msafam@\else
 \ifnum#2=\tw@\let\next@\msbfam@\fi\fi
 \mathchardef#1="#3\next@#4#5}
\def\mathhexbox@#1#2#3{\relax
 \ifmmode\mathpalette{}{\m@th\mathchar"#1#2#3}%
 \else\leavevmode\hbox{$\m@th\mathchar"#1#2#3$}\fi}
\def\hexnumber@#1{\ifcase#1 0\or 1\or 2
\or 3\or 4\or 5\or 6\or 7\or 8\or
 9\or A\or B\or C\or D\or E\or F\fi}

\font\tencmmib=cmmib10 
\font\sevencmmib=cmmib7 
 \skewchar\tencmmib'177 
\skewchar\sevencmmib'177 
\newfam\cmmibfam
\textfont\cmmibfam=\tencmmib
\scriptfont\cmmibfam=\sevencmmib 
\def\bpartial{\mathchar"0\hexnumber@\cmmibfam40}
\def\bn{\mathchar"0\hexnumber@\cmmibfam6E}
\def\bx{\mathchar"0\hexnumber@\cmmibfam78}
\font\tenmsa=msam10
\newfam\msafam
\textfont\msafam=\tenmsa
\edef\msafam@{\hexnumber@\msafam}
\mathchardef\dabar@"0\msafam@39

\def\lrcorner{\delimiter"5\msafam@79\msafam@79 }
\def\square{\mathchar"0\hexnumber@\msafam03}
\font\tenmsb=msbm10
\newfam\msbfam
\textfont\msbfam=\tenmsb
\edef\msbfam@{\hexnumber@\msbfam}
\def\Bbb#1{{\fam\msbfam\relax#1}}

\hsize=12.5  cm
\vsize=20.5 true cm

 1
\font\bbigrm=cmr12 scaled \magstephalf
\font\sc=cmcsc10 scaled \magstephalf
\font\srm=cmr9 
\font\sit=cmti9 

\font\frm=cmr8 
\font\fit=cmti8 

\advance\count0 by 1282 

\nopagenumbers
\if\pageno=1283 \footnote{ \centerline{(\pageno )}}\fi
\voffset=2\baselineskip
\headline={\ifnum\pageno=1283 Vol. {\bf 26} (1995)\hfill
{\sl ACTA PHYSICA POLONICA B}\hfill No {\bf 7}
\else{\ifodd\pageno\rightheadline\else\leftheadline\fi}\fi}
\def\rightheadline{\hfil {\it The Dirac Operator on Hypersurfaces}\hfil\folio}
\def\leftheadline{\rm\folio\hfil A. TRAUTMAN \hfil}

\topglue 1  true cm

\centerline{\bbigrm THE DIRAC OPERATOR ON 
HYPERSURFACES}\footnote{}
{\hsize=12.5  cm {\frm This paper is based, in part,
 on a lecture given at 
the}  
{\fit Workshop on Spinors, Twistors, and Conformal 
Invariants\/} {\frm at the Erwin Schr\"odinger Institute in 
Vienna (19--25 September 1994). During the} {\fit 
Workshop}, {\frm
there was a short ceremony honoring the memory of 
Professor Jan Rzewuski.}}
\footnote{}{\hsize=13.5 true cm {\frm
Research  supported in part by the Polish
Committee on Scientific Research (KBN) under Grant No. 2
P302 112 7.} }
\footnote{}{
\vskip3pt
\centerline{(1283)}}
\bigskip

\centerline{\sc Andrzej Trautman}
\bigskip

\centerline{\srm Instytut
 Fizyki Teoretycznej, Uniwersytet
Warszawski}
\centerline{\srm Ho\.za 69, 00-681 Warszawa, Poland}
\centerline{\srm and}
\centerline{\srm Laboratorio Interdisciplinare, 
Scuola Internazionale
Superiore di Studi Avanzati}
\centerline{\srm Via Beirut 2--4, 34014 Trieste, Italy}
\medskip

\centerline{\it Dedicated to the memory of Professor Jan Rzewuski}
\medskip

\centerline{\fit (Received March 15, 1995)}
\medskip

{\srm
{\narrower
 Odd-dimensional Riemannian spaces that
are non-orientable, but have a pin structure, require the
consideration of the twisted adjoint representation of the
corresponding pin group.  It is shown here how the Dirac
operator should be modified, also on even-dimensional
spaces, to make it equivariant with respect to the  action
of that group when the twisted adjoint representation is
used in the definition of the pin structure. An explicit
description of a  pin structure  on a hypersurface, defined
by its immersion in a Euclidean space, is used to derive a
{\sit Schr\"odinger transform\/} of the Dirac operator
 in that case. This is then
applied to obtain---in a simple manner---the spectrum
and eigenfunctions of the Dirac operator on spheres and real
projective spaces.\smallskip}}

{\srm PACS numbers 02.40+m, 03.65.Ge}
\bigskip

\centerline{\bf 1. Introduction}
\bigskip

Most of the research on the Dirac operator on Riemannian
spaces is restricted to the case of orientable manifolds.
It is of some interest to treat also the non-orientable
case that requires the introduction of pin structures. In
physics, even in the orientable case, one considers spinor
fields transforming under space and time reflections,
which are covered by elements of a suitable  pin group.
The generalization to the non-orientable case involves
interesting subtleties. First of all, for a real vector space
with a quadratic form of signature $(k,l)$, the Clifford
construction yields two groups, ${\Pin}_{k,l}$ and
${\Pin}_{l,k}$, which need not be isomorphic;  see [1]
 and Sec. 3.1 for a precise statement. This  fact
is of interest also to physics [2].
 There are
non-orientable spaces with a metric tensor field of
signature $(k,l)$ admitting either a ${\Pin}_{k,l}$-structure
or a $\Pin_{l,k}$-structure. If a space  admits a
$\Spin_{k,l}$-structure, then it is orientable and admits
both these structures. Real projective spaces and
quadrics provide the simplest examples of such situations
[3--5].
If the dimension $k+l$ is even, then one
can use either the adjoint or the twisted adjoint
representation  of ${\Pin}_{k,l}$. If one uses the
twisted adjoint representation, as one has to do when
$k+l$ is odd, then   the classical Dirac operator  (see, e.g.,
[6--8]) needs to be modified to make it
equivariant with respect to the action of the pin group
[5,9].
In this paper, the relation between the adjoint and the
twisted adjoint representation of the pin group  is
considered  in some detail (Section 3). In Section 4, the definition of 
spin and pin structures is illustrated on the example of spheres and
real projective spaces. The
form of the modified Dirac operator is recalled in
Section 5. A canonical pin structure on a hypersurface
immersed in a Euclidean space is described in Section 6
and shown to have a trivial associated bundle of `Dirac' or
`Pauli' spinors . A convenient formula for the 
`Schr\"odinger transform' of the modified Dirac
operator on such hypersurfaces is derived in Section 7. As
an  illustration, the spectrum and the eigenfunctions
of the Dirac operator  on real projective spaces are found
on the basis of the corresponding results for spheres 
(Section 8).
\bigskip
\centerline{\bf 2. Notation and Preliminaries}
\bigskip
This paper is a continuation of [5] and [9]; it uses the 
notation and terminology introduced there. To make the 
paper self-contained, some of the notation is summarized below.
\bigskip
\centerline{\it 2.1 Clifford algebras and pin groups}
\bigskip
Throughout this paper, by an algebra I mean an associative
algebra with a unit element. A
homomorphism of algebras is understood to map one unit
into another.  A representation of an algebra is a 
homomorphism of the algebra into the algebra $\End S$ of 
all endomorphisms of a vector space $S$.  If $V$ is a 
finite-dimensional vector space,
then $V^*$ denotes its dual and the value $f(v)$ of the
1-form $f\in V^*$ on $v\in V$ is often denoted by $\langle
v, f\rangle $. If $h:V\to W$ is a linear map (homomorphism
of vector spaces), then its {\it transpose\/}  ${^t}h:W^*
\to V^*$ is defined by $\langle v, {^t}h(f)\rangle =\langle
h(v),f\rangle$ for every $v\in V$ and $f\in W^*$.  Let $V$
be a real,  $m$-dimensional vector space with an
isomorphism $h:V\to V^*$ which is symmetric, $h={^t}h$,
and such that the quadratic form  $V\to \Bbb R$, given by
$v\mapsto \langle v, h(v)\rangle $ is of signature $(k,l),\;
k+l=m$. One says that the pair $(V,h)$ is a {\it quadratic
space} of dimension $m$ and signature $(k,l)$. The
corresponding {\it Clifford algebra\/} (see, e.g., [1,8,10])
 $$
  {\Cl}(h)={\Cl}^0 (h)
\oplus {\Cl}^1 (h), 
$$
 contains $\Bbb R \oplus V$ and
is $\Bbb Z _2$-graded by the {\it main automorphism\/}
$\alpha$ characterized by $\alpha (1)=1$ and $\alpha
(v)=-v$ for every $v\in V$. Every $a\in {\Cl}(h)$ is
decomposed into its even and odd components, $a_0$ and
$a_1$, respectively, such that $a_\varepsilon \in
{\Cl}^\varepsilon (h)$ and $a=a_0+a_1 =\alpha (a_0
-a_1 )$. For every $v\in V$, its Clifford square is $v^2
=\langle v, h(v)\rangle$. Assume $V$ to be oriented and
let  $(e_1,\ldots ,e_m )$ be an orthonormal frame in $V$
of the preferred orientation.  The square of the volume element
${\vol}(h) =e_1 \ldots e_m$ is 
$$
{\vol}(h)^2 =i(h)^2 ,\quad{\rm where}\quad i
(h)\in \{1,{\ri}\}. 
$$
For every $v\in V$ one has
$\vol (h)\vts  v=(-1)^{m+1}v\vts  {\vol} (h)$.
Therefore,  if $m$ is even, then $\alpha$ is an inner
automorphism, $\alpha (a)= \vol (h)\vts   a
\vts  \vol (h) ^{-1}$.

It follows from the universality of Clifford algebras that
the Clifford map
 $$ V\to \Cl (h), \quad v\mapsto
\vol (h) v, $$ 
extends to the homomorphism of
algebras, 
$$ j :\Cl
((-1)^{m+1}i(h)^2 h)\to \Cl (h),
\eqno{(1)} $$ 
such that  
 $j (1)=1$ and $j (v)=
\vol (h) v $ for $ v\in V$.
For $m$
even, this homomorphism is bijective and respects the $\Bbb
Z_2$-grading of the algebras. If $m$ is even and
$\vol (h) ^2 =1$, then $j  :\Cl (-h)\to
\Cl (h)$ is an isomorphism of algebras. If $m$ is
even and $\vol (h) ^2 =-1$, then the algebras
$\Cl (h)$ and ${\Cl}(-h)$ are not isomorphic
and $j$ is an inner automorphism of $\Cl (h)$ given
by $j (a)=\textstyle {1\over 2} (1+{\vol} (h)
)a(1-{\vol} (h) )$. If $m$ is odd, then the
homomorphism (1) is onto the even subalgebra
${\Cl}^0 (h)$. In this case, the volume element
corresponding to $\vol (h) ^2 h$ has a
positive square. Therefore, if $m$ is odd and $h$ is such
that $\vol (h) ^2 =1$, then $j(\vol
(h))=1$ and there is the exact sequence of
homomorphisms of algebras, 
$$ 0\to {\Cl}
^- (h)\to {\Cl} (h)\buildrel{j}\over {\to}{\Cl} ^0
(h)\to 0, $$
 where ${\Cl}^- (h)=\{a\in {\Cl}
(h):{\vol}(h) a=-a\}$ is the subalgebra of
anti-selfdual elements of $ \Cl (h)$. There is no
analogous sequence for $m$ odd and $h$ such that
$\vol (h) ^2 =-1$.
 For $m$ odd, the algebras
$\Cl (h)$ and ${\Cl}(-h)$ are never
isomorphic. The algebras ${\Cl} ^0 (h)$ and
${\Cl} ^0 (-h)$ are isomorphic irrespective of $m$
and $h$.
An element $u\in V$ is said to be a {\it unit vector\/} if
either $u^2 =1$ or $u^2 =-1$. The group ${\Pin}(h)$
is defined as the subset of ${\Cl}(h)$ consisting of
products of all finite sequences of unit vectors; the group
multiplication is induced by the Clifford
product.\footnote{$^1$}{\frm
 This definition of the pin group follows [5,8,11]
 and can be traced to Cartan, see Sections 12, 97
and   127 in [12].
 An equivalent definition, using the
notion of spinor norm, and based on the (semi-)simplicity of
the Clifford algebras, is in [13,14].}
The spin group
is  
${\Spin}(h)={\Pin}(h) \cap {\Cl}^0
(h)$.
The Lie algebra ${\rm spin\vts  }(h)$ of
${\Spin}(h)$ can be identified with the subspace of
${\Cl}^0 (h)$ spanned by all elements of the form
$uv-vu$, where $u,v\in V$. The Lie bracket in
${\rm spin\vts  }(h)$ coincides with the
commutator induced by the Clifford product.

If $V=\Bbb R ^{k+l}$ and one wants to specify the
signature $(k,l)$ of $h$, then one writes
${\vol}_{k,l}$, $\;{\Cl}_{k,l}$, $\;
{\Pin}_{k,l}$   and ${\Spin} _{k,l}$
instead of ${\vol}(h)$, $\;{\Cl}(h)$,$\;
{\Pin}(h)$ and ${\Spin}(h)$,
respectively; a similar notation is used for the orthogonal
groups ${\Ort} (h)$ and ${\SO}(h)$. Since
the groups ${\Spin}(h)$  and ${\Spin}
(-h)$ are isomorphic, one writes ${\Spin} _m$
instead of ${\Spin}_{m,0} = {\Spin}
_{0,m}$.  Since ${\vol}_{2n,0}^2
={\vol}_{0,2n}^2$ one can also write
${\vol}_{2n}$ instead of
${\vol}_{2n,0}$ or ${\vol}_{0,2n}$.
\bigskip
\centerline{\it 2.2 Notation concerning smooth manifolds and
bundles}
\bigskip 
 All
manifolds, maps and bundles are assumed to be smooth;
manifolds are paracompact and bundles are locally trivial.
If $\pi : E\to M$ and $\sigma :F\to N$ are two bundles , then
the pair $(f,f')$ of maps $f:M\to N$ and $f':E\to F$  is a {\it
morphism of bundles\/} if $\sigma \circ f'=f\circ \pi$. A
bundle is trivial if it is isomorphic to a Cartesian product
of its base by the typical fiber. A map $s:M\to E$ is a {\it
section\/} of $\pi$ if $\pi \circ s={\vts \id}_M$. For
every manifold $M$, there is the {\it tangent bundle\/}
$TM\to M$. If $f:M\to N$ is a map of manifolds,
then $Tf :TM\to TN$ is the derived map of their tangent
bundles and $(f,Tf)$ is a morphism of bundles. For $x\in M$,
there is the linear map $T_x f:T_x M\to T_{f(x)}N$ of the
fiber $T_x M$ of the bundle $TM\to M$ into the
corresponding fiber of the other bundle. Given a bundle
$\sigma :F\to N$ and a map  $f:M\to N$, one defines the
bundle $\pi :E\to M$ {\it induced\/} by $f$ from $\sigma$
as follows: $E=\{(x,q)\in M\times F:\sigma (q)=f(x)\}$ and
$\pi (x,q)=x$. There is then also a canonical map, $f':E\to
F$, given by $f'(x,q)= q$ and the pair $(f,f')$ is a morphism
of bundles. A {\it Riemannian space\/} is a connected
manifold $M$ with a metric tensor field $g$ which need not
be definite; if it is, then one refers to $M$ as a {\it
proper\/} Riemannian space.  For every $x\in M$, the
metric tensor defines a symmetric isomorphism $g_x :T_x
M\to {T_x}^* M$. If $M$ is a Riemannian space, then there is
a quadratic space $(V,h)$ such that, for every $x\in M$,
there is a linear isometry $p:V\to T_x M$, i.e. a linear
isomorphism such that ${^t}p\circ g_x \circ p=h$. One says
that $(V,h)$ is {\it local model\/} of the Riemannian
space and that $p$ is an {\it orthonormal frame at\/} $x$.
If $(e_\mu)$ is an orthonormal frame in $V $, then $p$ can
be identified with the collection of vectors $(p_\mu )$,
where $p_\mu =p(e_\mu ),\;\;\mu =1,\dots ,m =\dim V
=\dim M$.

If $\omega$ is a differential form on a manifold, then
$\rd \vts  \omega$ is its exterior derivative. Wedge
denotes the exterior product of forms. If $X$ is a vector
field on $M$ and $\omega$ is a $(p+1)$-form, then
$X\vts  \lrcorner\ts  \omega$ is the $p$-form such that
$(X\vts  \lrcorner\ts  \omega )(X_1 ,\dots ,X_p )=\omega (X,X_1
,\dots ,X_p )$ for every collection $(X_1 ,\dots ,X_p )$ of
vector fields on $M$. In particular, if $f:M\to \Bbb R$ and
$X$ is a vector field, then $X\vts  \lrcorner\ts  \rd f=
\langle X, \rd f\rangle =X(f)$ is the
derivative of the function $f$ in the direction of the
vector field $X$.

By a group is meant here a Lie group; a subgroup is a
closed Lie subgroup. An exact sequence
of group homomorphisms
$1\to K\buildrel
{k}\over{\to}G\buildrel{l}\over{\to}H\to 1$
is said to define $G$ as an {\it extension of $H$ by $K$}.
Two extensions,
$K\buildrel{k}\over{\to}G\buildrel{l}\over{\to}H$ and
$K\buildrel{k'}\over{\to}G'\buildrel{l'}\over{\to}H$,
 of the group $H$
by the group $K$, are {\it equivalent\/} if there is an
isomorphism of groups $f:G\to G'$ such that $f\circ k=k'$
and $l'\circ f=l$.
Given a representation $\gamma :G\to  {\GL}(S)$ of the 
group $G$ in a vector space $S$ and a 
homomorphism $\iota :G\to G'$ of groups, one
says that a representation $\gamma ':G'\to {\GL}(S)$
{\it extends\/} $\gamma$ (relative to $\iota$) if
$\gamma '\circ \iota =\gamma$.
\smallskip
 A {\it
principal bundle\/} with structure group $G$ (`principal
$G$-bundle') and projection $\pi$ of its total space $P$ to
the base manifold $M$ is sometimes represented,
symbolically, by the sequence $G\to P\buildrel{\pi}\over{\to}
M$. The group $G$ is assumed to act on $P$ to the right:
there is a map $\delta :P\times G\to P$ such that, if
$\delta (a) (p)=\delta (p,a)$, then $\pi \circ \delta
(a)=\pi$,  $\; \delta (a) \circ \delta (b) =\delta (ba)$ and
$\delta (1_G )={\id}_P$, where $p\in P,\; a,b\in
G$ and $1_G$ is the unit of $G$. One  writes $pa$ instead of
$\delta (p,a)$. A principal bundle admitting a section $f$ is
trivial, i.e. isomorphic (in the category of principal
bundles) to the product bundle $M\times G\to M$; a {\it
trivializing map\/} (isomorphism of principal bundles) is
given by $(x,a)\mapsto f(x)a$, where $x\in M$ and $a\in G$.
Let there be given a left action of the group $G$ on the
manifold $S$, i..e. a map $\gamma :G\times S\to S$ such
that, if $\gamma (a)(\varphi)=\gamma (a,\varphi )$, then
$\gamma (a) \circ \gamma (b) =\gamma (ab)$ and $\gamma
(1_G )={\id}_S$ for every $a,b\in G$ and $\varphi
\in S$. One then defines the bundle $\pi _E :E\to M$, {\it
associated\/} with $P$ by $\gamma $. Its typical fiber is
$S$ and its total space $E$, often denoted by $P\times
_\gamma  S$, is the set of all equivalence classes of the
form $[(p,\varphi )]$, where $(p,\varphi )\in P\times S$ and
$[(p',\varphi ')] =[(p,\varphi )]$ if, and only if, there exists
$a\in G$ such that $p'=pa$ and $\varphi =\gamma
(a)\varphi '$. The projection is given by $\pi _E
([(p,\varphi )])=\pi (p)$. If $S$ is a vector space, then the
associated bundle is a {\it vector bundle}. A
homomorphism $\iota :G\to G'$ of groups defines a left
action of $G$ on $G'$, viz. $(a,b)\mapsto \iota (a)b$, where
$a\in G$ and $b\in G'$; the corresponding  bundle $P\times
_\iota G'\to M$, associated with $P\to M$, is a principal
$G'$-bundle.
\eject
\centerline{\bf 3. Representations of the pin groups}
\bigskip
\centerline{\it 3.1 The vector representations}
\bigskip
For every invertible  $u\in V$, the map $v\mapsto
-uvu^{-1}$ is a reflection in the hyperplane orthogonal to
the vector $u$; this observation leads to the definition of
the {\it twisted adjoint\/} vector representation $\rho$
of the group $\Pin (h)$ in $V$: for every $a\in
\Pin (h)$ the map $\rho (a):V\to V$, given by
$$ \rho (a)\vts v=\alpha (a)\vts  v\vts  a^{-1}
\eqno{(2)} $$ is orthogonal,
$$ {^t}\rho (a)\circ h \circ \rho (a)=h,
\eqno{(3)} $$ and there is the exact sequence
of group homomorphisms $$ 1\to \{1,-1\} \to
\Pin (h)\buildrel{\rho}\over{ \to} {\Ort}(h)\to 1. $$
 Replacing in (2) the vector $v$ by
the $\mu$th vector $e_\mu$ of an orthonormal frame in
$V$, one obtains $$ e_\nu {\rho ^\nu}_\mu
(a)=\alpha (a)e_\mu a^{-1} \eqno{(4)} $$
In this equation, and elsewhere in this paper, there is
tacitly assumed a summation (the {\it Einstein
convention\/}) over the range of tensor indices
appearing in contragredient pairs.

The {\it adjoint\/} vector representation Ad is defined
by $$ {\Ad}(a)\vts  v=a\vts  v\vts  a^{-1} $$ and leads to the
exact sequences of group homomorphisms 
$$1\to\left \{\matrix{\{1,-1\}\cr
\{ 1,-1,{\vol} (h),
-{\vol} (h)\}\cr}\right\}\to\Pin (h)\buildrel{\Ad}\over
{\lto}
\left\{\matrix{\Ort (h)\cr
\SO (h)\cr}\right\}\to 1 
\;\cases{{\rm for}\;\; m\!& even,\cr
{\rm for}\;\; m\!& odd.\cr}
$$
 The
homomorphisms $\rho$ and $\Ad$ coincide
when restricted to $\Spin (h)$. For every
quadratic space $(V,h)$, irrespective of the parity of $m$,
there is the exact sequence
 $$ 1\to \Bbb
Z_2  \to \Spin (h) \buildrel{\rho}\over{\to}
\SO (h)\to 1,\eqno{(5)} $$
 where
$\Bbb Z_2 =\{1,-1\}$.

For every  {\it even\/}-dimensional quadratic  space
$(V,h)$,  one can consider four central extensions of
$\Ort (h)$ by $\Bbb Z_2 $, associated with the
groups $\Pin (\pm h)$, namely
 $$ \rho \;\; {\rm and}\;\;
\Ad :\Pin (h)\to \Ort
(h), \;\;{\rm and}\;\; \rho\;\; {\rm and}\;\; \Ad
:\Pin (-h)\to \Ort (h), $$
 but, in each
case, only two among the four are inequivalent.
  Indeed, if $m$ is even, then 
  $$ \rho
=\Ad\circ j \eqno{(6)} $$
 as may be seen by checking
that both sides of (6) coincide on the generating subset $V$.
More precisely:
\smallskip
(i) if $\vol (h)^2 =1$, then the extensions

$$ \Bbb Z_2 \to \Pin (\pm h)
\buildrel{\Ad}\over{\lto }\Ort (h) $$ 
are
equivalent to the corresponding  extensions
 $$ \Bbb Z_2
\to \Pin (\mp h)\buildrel{\rho}\over{\to}
\Ort (h); $$

(ii) if $\vol (h)^2  =-1$, then the extensions 
$$
\Bbb Z_2 \to \Pin (\pm h)
\buildrel{\Ad}\over{\lto }\Ort (h) $$
 are
equivalent to the corresponding extensions $$ \Bbb Z_2
\to \Pin (\pm h)\buildrel{\rho}\over{\to}
\Ort (h). $$

To summarize, we have
\proclaim Proposition 1. 
 For every real
quadratic space $(V,h)$,  there are two inequivalent
central extensions of $\Ort (h)$  by $\Bbb Z_2$,
given by $$ \Bbb Z_2 \to \Pin (
h)\buildrel{\rho}\over{\to} \Ort (h)\;\;{\rm  and}\;\;
\Bbb Z_2 \to \Pin (-h)\buildrel{\rho}\over{\to}
\Ort (h),\eqno{(7)} $$
 where $\rho$
is as in {\rm (2)}. By restriction to
$\Spin (h)$  each of these extensions reduces
to the one given by {\rm (5)}.\par
  
Note that
for $k=l$ (neutral signature) the groups
$\Pin (h)$ and $\Pin(-h)$ are
isomorphic, but the extensions (7) are not. There
are also extensions of $\Ort(h)$ by $\Bbb Z_2$
that do not come from the Clifford construction [15].
The (untwisted) adjoint representation seems to be the
first to have attracted attention. It has been much used
by physicists in the theory of the Dirac equation of the
electron; see, e.g., [6,16].
The twisted representation is implicit
in \'E. Cartan's approach to spinors, see \S 58 and \S 97 in
[12].
Explicitly, it has been defined by Atiyah {\it et
al.\/} in [17].
It follows from the preceding remarks
that, for even-dimensional spaces, one can use either of
the two representations, but in the case of odd
dimensions, only $\rho$ provides a cover of the full
orthogonal group. For this reason and for uniformity, from
now on, only $\rho$ is used in the definition of pin
structures.
\bigskip
\centerline{\it 3.2 The spinor representations}
\bigskip
In this paper, by a {\it spinor representation\/} of a
group $\Pin (h)$ or $\Spin (h)$ is
understood a representation obtained by restriction, to
the group,  of a representation of the algebra
$\Cl (h)$ in a finite-dimensional {\it
complex\/} vector space $S$, the space of {\it spinors}. If
$\gamma :\Cl (h)\to \End S$ is any
representation of the algebra, then the group
representation, obtained by restriction to
$\Pin (h)$, is denoted by the same letter
$\gamma$; similar abuses of notation and terminology are
made throughout the paper.  Given an orthonormal
frame $(e_\mu )$ in $V$, one defines the `Dirac matrices'
(automorphisms of $S$) by $\gamma _\mu =\gamma
 (e_\mu )$. 
The following Proposition summarizes well known facts 
about complex representations of
 real Clifford algebras [5,8,10,17].
\proclaim Proposition 2. Let $(V,h)$ be a quadratic space 
of dimension $m$ and let $\nu$
 denote a positive integer.\hfil\break
(i) If $m$ is {\rm even}, $m=2\nu$, then
the algebra $\Cl (h)$ is central simple and, as
such, has only one, up to equivalence, faithful and
irreducible {\rm Dirac representation\/} $\gamma$ in a
vector space $S$, which turns out to be of complex
dimension $2^\nu$. The  restriction of $\gamma$ to
$\Cl ^0 (h)$ decomposes into the direct sum
$\gamma _+ \oplus \gamma _-$ of two
complex-inequivalent {\rm Weyl representations}.
 In a notation
adapted to the decomposition $S=S_+\oplus S_-$  of the space of Dirac spinors 
into the direct sum of the spaces $S_+$ and $S_-$ of Weyl spinors,
 the Dirac matrices
are of the form
 $$ \gamma _\mu =\left ( \matrix{0
&\gamma ^- _\mu\cr
 \gamma ^+ _\mu &0\cr}\right ). $$
(ii) If $m$ is {\rm odd}, $m=2\nu -1$, then the algebra
$\Cl ^0 (h)$ is central simple and has a faithful
and irreducible {\rm Pauli representation\/} in a space 
of complex dimension $2^{\nu -1}$. This representation
extends to two representations, $\sigma$ and $\sigma
\circ \alpha$, of the full algebra $\Cl (h)$ in the same space,
by putting $\sigma (\vol (h))= i (h)\vts  
\id$. These  representation, also referred
to as Pauli representations of $\Cl (h)$, are
complex-inequivalent and irreducible, but  faithful only
when $i (h)={\ri}$.  A faithful, but reducible, {\rm
Cartan representation} $\gamma$ of $\Cl (h)$
 is defined as $\gamma = \sigma \oplus
(\sigma \circ \alpha )$. Therefore, if $\sigma _\mu
=\sigma (e_\mu )$, then
 $$ \gamma _\mu =\left (
\matrix{\sigma _\mu &0\cr 0& -\sigma _\mu
} \right ). $$
The commutant of the Cartan representation $\gamma$ is 
generated by $\gamma (\vol (h))$.
 The elements of the carrier spaces of the representations 
 $\gamma$ and $\sigma$ are now
called Cartan and Pauli spinors, respectively.\par

 The names
of Dirac, Weyl and Pauli are used by physicists mainly in
connection with spinors associated with vector spaces of
low dimension. In mathematics, the Weyl representations
$\gamma _\pm$ are usually denoted by $\Delta ^\pm$ 
 and sometimes referred to as half-spinor
representations [8,13]. The Cartan representation seldom
appears because it is decomposable. In this paper, I
identify the representations by using one of the above
names; thus the letters $\gamma$ and $\gamma'$ can
 denote any one of the spinor
representations, depending on the context.

If $\gamma :\Cl (h)\to \End S$ is as in
 Prop. 2, then the {\it helicity\/}
automorphism of the representation $\gamma$ in $S$ is
 $\gamma (\vol (h))=\gamma _1 \dots \gamma_m$
so that Weyl (resp., Pauli) spinors are its eigenvectors
for $m$ even (resp., odd). The foregoing remarks can be
supplemented by
\proclaim Proposition 3.
 Let 
$(V,h)$ be a quadratic space of dimension
 $m=2\nu\; {\rm (}$resp., $2\nu
-1{\rm )}$. There is a faithful representation $\gamma $
of the Clifford algebra $\Cl (h)$ in a complex
vector space $S$ of dimension $2^\nu$ such that the
representations $\gamma$ and $\gamma \circ \alpha$ are
complex-equivalent. The representation is unique, up to
complex equivalence, and irreducible ${\rm (}$resp.,
decomposable into two irreducibles${\rm )}$. By
restriction to the even subalgebra $\Cl ^0 (h)$,
the representation $\gamma $ decomposes into the direct
sum of two irreducible representations, each defined in a
complex space of dimension $2^{\nu -1}$.  The
isomorphism $\gamma _{m+1}$ intertwining the
representations $\gamma$ and $ \gamma \circ \alpha$ can
be taken  to act on the Dirac ${\rm (}$resp., Cartan${\rm )}$
spinor $(\varphi ,\psi )$ so that $\gamma
_{m+1}(\varphi ,\psi ) =({\ri}\varphi ,
-{\ri}\psi )$ ${\rm (}$resp.,  $\gamma _{m+1}(\varphi
,\psi )=(-\psi , \varphi ) {\rm )}$. Irrespective of the
parity of $m$, one has 
$$ \gamma _{m+1}^2
=-\ts \id_S\quad{\rm and}\quad \gamma
_{m+1}\gamma _\mu +\gamma _\mu \gamma
_{m+1}=0,\eqno{(8)} $$ for $\mu =1,\dots ,m$. 
\par
The intertwining isomorphism is not unique; 
see [5] for a precise 
statement on the `Dirac intertwiner' ${\ri}\gamma_{m+1}$.  
By applying  a spinor representation $\gamma$ to both
sides of (4), one obtains
 $$
\gamma _\nu \vts  {\rho ^\nu}_\mu (a)=\gamma \circ \alpha
(a)\vts  \gamma _\mu\vts   \gamma (a).\eqno{(9)}
$$
\bigskip
 \centerline{\it 3.3 Extension of a spinor
representation from dimension $m$ to $m+1$}
\bigskip
For every pair $(k,l)$ of non-negative integers, there is
the {\it isomorphism\/} of algebras, 
$$
 \iota
:\Cl _{k,l}\to\Cl ^0
_{k,l+1}\quad{\rm given}\;{\rm by}\quad\iota (a_0 +a_1 )=a_0
+a_1 e_{k+l+1}.
 $$
\proclaim Proposition 4. If $\gamma'$ is a representation 
of $\Cl _{k,l+1}$ in a complex vector space (of spinors), 
then $\gamma =\gamma'\circ\iota$ 
is a representation of $\Cl_{k,l}$ in the same  space. In particular:\break
(i) If $k+l$ is even and $\gamma'$ is a Pauli 
representation, then $\gamma$ is the Dirac representation.\break
(ii) If $k+l$ is odd and $\gamma'$ is the Dirac 
representation, then $\gamma$ is the Cartan 
representation. Moreover, if $\gamma'_\pm$ are the Weyl 
components of $\gamma'$, i.e. $\gamma' |\Cl^0 
_{k,l+1}=\gamma'_+ \oplus \gamma'_-$, then $\gamma_\pm 
=\gamma'_\pm \circ \iota$ are the Pauli components of 
$\gamma$, i.e. $\gamma=\gamma_+ \oplus\gamma_-$.
\par
\noindent {\sl Proof.} Since $\iota$ is injective, if $\gamma'$ is 
faithful, then so is $\gamma$. In case (i), the Pauli representation 
$\gamma'$ is faithful unless $\vol ^2 _{k,l+1}=1$. If $\vol ^2 
_{k,l+1}=1$, then the kernel of $\gamma'$ is either the subalgebra $\Cl^+ 
_{k,l+1}$ of selfdual elements or the subalgebra $\Cl^- 
_{k,l+1}$ of anti-selfdual elements of $\Cl_{k,l+1}$, see Sec. 2.1. Since 
$\Cl^\pm _{k,l+1}\cap \Cl^0 _{k,l+1}=\{0\}$, the representation $\gamma= 
\gamma'\circ\iota$ is faithful in every case when $m=k+l$ is even. In 
case (ii), the Dirac representation $\gamma'$ of $\Cl_{k,l+1}$ in $S$
is faithful. Therefore, the representation $\gamma$
 is also faithful. Let $\gamma_i =\gamma'(e_i ),\;i=1,\dots ,m+1=k+l+1$, be the 
Dirac matrices. Then $\gamma\circ\alpha (a)= \gamma_{m+1}
\gamma (a)\gamma^{-1}_{m+1}$ for every $a\in \Cl_{k,l}$ and, 
by Prop. 3, $\gamma$ is the Cartan representation. 
Since $\gamma (e_i ) =\gamma' (e_i e_{m+1})=\gamma_i\vts 
\gamma_{m+1}$ for $i=1,\dots ,m$, the 
helicity automorphisms of $\gamma'$ and $\gamma=\gamma'\circ\iota$ 
are equal. Therefore, 
the decompositions of $S$ into spaces of Weyl and Pauli spinors  coincide.
\hfill $\square$
\smallskip
 By iteration of the above,  one can
obtain, for $k+l$ {\it odd}, two Pauli representations  of
$\Cl_{k,l+2}$ extending the Cartan
representation  of $\Cl _{k,l}$. Similarly, for
$k+l$ {\it even}, there are two Weyl representations of
$\Cl^0 _{k,l+2}$ extending the Dirac
representation of $\Cl _{k,l}$. One cannot,
however, go beyond that without changing the dimension
of the space of spinors underlying the representations.

  By restriction, the isomorphism of algebras $\iota$  gives rise to the {\it
monomorphism\/} of groups
 $$
  \iota
:\Pin _{k,l}\to\Spin 
_{k,l+1}.\eqno{(10)}
 $$ 
 The corresponding
monomorphism of the (pseudo-)orthogonal groups,
$$
 \kappa :\Ort _{k,l}\to
\SO _{k,l+1}\;\; {\rm  such}\;{\rm that }\;\;\kappa \circ \rho
=\rho \circ \iota, \eqno{(11)}
 $$
  satisfies
$\kappa (A)\vts  e_\mu =A\vts  e_\mu ,\;\; \mu =1,\dots  ,k+l$, and
$\kappa (A)\vts  e_{k+l+1}=({\det }\vts   A)\vts  e_{k+l+1}$ for
every $A\in \Ort _{k,l} $.

By restricting the representations $\gamma'$ and $\gamma$,
 referred to in Prop. 4, to 
the groups $\Pin_{k,l+1}$  and $\Pin_{k,l}$ one obtains  
representations of pin and spin groups; there are statements about 
extensions of spinor representations of these groups analogous  to those 
appearing in the Proposition.
\bigskip
\centerline{\bf 4. Pin structures and bundles of spinors}
\bigskip
\centerline{\it 4.1 Definitions}
\bigskip

Let $(V,h)$ be a local model of an $m$-dimensional
Riemannian manifold $M$ and let $\pi : P
\to M$ be the bundle of  all orthonormal frames of $M$. A
$\Pin (h)$-{\it structure\/} on $M$ is a  principal
$\Pin (h)$-bundle $\varpi : Q\to M$, together
with a morphism $\chi :Q\to P$ of principal bundles over
$M$ associated with the epimorphism  $\rho
:\Pin (h)\to \Ort (h)$. The morphism
condition means that $\varpi =\pi \circ \chi $ and that, 
for every $q\in Q$ and $a\in \Pin (h)$, one has $\chi (qa)= 
\chi (q)\rho (a)$.
The expression $\Pin _{k,l}\vts$-{\it structure\/} is used when
one wants the signature of $h$ to appear explicitly. For
brevity, we shall describe a $\Pin (h)$-structure
 by the sequence
$$ \Pin (h) \to Q
\buildrel{\chi}\over{\to} P \buildrel{\pi}\over{\to} M.
 \eqno{(13)}
$$
Another pin 
structure over the same manifold $M$, $\Pin (h) \to Q'
\buildrel{\chi'}\over{\to} P \buildrel{\pi}\over{\to} M$ 
 is said to be {\it equivalent\/} to the structure (13) if there is a 
diffeomorphism $f:Q\to Q'$ such that $\chi'\circ f=\chi$ and 
$f(qa)=f(q)a$ for every $q\in Q$ and $a \in \Pin(h)$.

If $M$ is orientable and admits a $\Pin (h)$-structure, then
it  has a spin structure. In an abbreviated style, similar to
that of (13), it may be described  by the
sequence of maps 
$$ \Spin (h) \to
SQ \to SP \to M, \eqno{(14)} $$
 where $SP$
is now an $\SO (h)$-bundle. One often abbreviates the 
expression `$M$ has a spin structure' to `$M$ is spin'. Equivalence of 
spin structures is defined similarly to that of pin structures.

 Let $M$ be a
Riemannian space with a  $\Pin (h)$-structure
(13) and let $\gamma$ be a spinor
representation of the group $\Pin  (h)$ in $S$,
as described in Prop. 2. The complex vector
bundle $\pi _E :E\to M$, with typical fiber $S$, associated
with $Q$ by $\gamma$, is the {\it bundle of spinors of
type\/} $\gamma$. If the dimension $m$ of $M$ is even
(resp., odd),  then $E$ is called a bundle of Dirac (resp.,
Cartan) spinors. For $m$ odd, $m=2\nu -1$, one can 
also take the
representation $\sigma :\Pin (h)\to
\GL(2^{\nu -1},\Bbb C)$  to define the bundle of Pauli spinors
over $M$. Similarly, if $m$ is even and $M$ has a spin
structure, then there are two bundles of Weyl spinors
over $M$. 

Let $M$ be a
Riemannian manifold with a pin structure (13). A
{\it spinor field\/} (some authors say: a `pinor'
field) of type $\gamma$ on $M$ is a section of $\pi _E$.
The (vector) space of such sections is known to be in a
natural and bijective correspondence with the set of all
maps $\psi :Q\to S$ equivariant with respect to the action
of $\Pin (h),\;\;
\psi \circ \delta
(a)=\gamma (a^{-1})\psi ,  $
for every $a\in\Pin (h)$. It is convenient to
refer to $\psi$ itself as a spinor field  of type $\gamma$
on $M$. Depending on whether $E$ is a bundle of Dirac,
Weyl, Cartan or Pauli spinors, one refers to its sections as
Dirac, Weyl, Cartan or Pauli spinor fields, respectively.

The existence of a pin (or spin) structure on a Riemannian 
manifold $M$ imposes topological conditions on $M$. They 
are expressed in terms of the Stiefel-Whitney classes $w_i\in
H^i (M,\Bbb Z_2 )$
associated with the 
tangent bundle of the manifold; see, e.g. [1]
and the applications of the Karoubi theorem given in 
[3--5].
\bigskip
\centerline {\it 4.2 Remarks on the triviality of associated
bundles}
\bigskip
\proclaim Proposition 5. 
 The vector bundle $E\to M$, associated with
the principal $G$-bundle $Q\to M$ by a representation
$\gamma$ of $G$ in $S$, is trivial if, and only if, there
exists a group $G'$, a homomorphism $\iota :G\to G'$, and
an extension  $\gamma ':G'\to {\GL }(S)$  of
$\gamma$, such that the associated principal $G'$-bundle
$Q\times_\iota G' \to M$ is trivial. 
\par
 \noindent {\it Proof.} Indeed, if $E$ is trivial as a vector bundle,
then there is a trivializing map $E\to M\times S$,
$\;[(q,\varphi )]\mapsto (\pi (q),g(q)\varphi )$, such that
$g:Q\to {\GL }\vts  (S)$ and $g(qa)=g(q)\circ \gamma
(a)$ for every $q\in Q,\; \varphi \in S$ and $a\in G$. Taking
$G'={\GL }\vts  (S)$ and $\iota=\gamma$ one sees that
$\gamma ' =\id $ extends $\gamma$. The
principal bundle $Q\times _\iota G'\to M$ is trivial
because it has a global section corresponding to the
equivariant map $e:Q\to G'$, where $e(q)=\iota (q)^{-1}$
for every $q\in Q$. Conversely, given an extension
$\gamma '$ of $\gamma$ and a homomorphism $\iota :G\to
G'$ such that $Q\times _\iota G'\to M$ is trivial, there is a
map $e:Q\to G'$ such that $e(qa)=\iota (a^{-1})e(q)$ for
every $q\in Q$ and  $a\in G$. If $g:Q\to {\GL }\vts  (S)$
is given by $g(q)=\gamma '(e(q)^{-1})$, then the map
$[(q,\varphi )]\mapsto (\pi (q),g(q)\varphi)$, which is
well-defined because of $\gamma '\circ \iota =\gamma$,
trivializes the vector bundle $E\to M$.{\hfill  $\square$ }
\smallskip
 If $G$ is a
subgroup of $G'$, then there is the principal $G$-bundle
$\pi :G'\to G'/G$. The action of $G$ on $G'$ given by the left
translations defines the associated  principal $G'$-bundle
$G'\times _\gamma G' \to G'/G$ that is trivial: a
trivializing map is given by $[(a,b)]\mapsto (\pi (a),ab)$,
where $a,b\in G'$.
\smallskip
\noindent{\sl Corollary.}  If there is  a representation $\gamma '$ of
$G'$ in $S$ extending the representation $\gamma :G\to
\GL (S)$, then the bundle $G'\times _\gamma S
\to G'/G$, associated with $\pi :G'\to G'/G$ by $\gamma$, is
trivial. 
\smallskip
 Indeed,   a
trivializing isomorphism is given by $[(a,\varphi
)]\mapsto (\pi (a),\gamma '(a)\varphi )$, where $a\in G'$
and $\varphi \in S$.
\bigskip
\centerline{\it 4.3 Examples} 
\bigskip
{\sl A. The spheres.} For
every $m>1$, the unit sphere $\Bbb S _m \subset \Bbb R ^{m+1}$
 has a unique spin structure described by

$$
 \Spin _m \to
\Spin _{m+1} \to \SO _{m+1}\to \Bbb S _m .
\eqno{(15)}
$$
 Its bundle of Dirac
($m$ even) or Pauli ($m$ odd) spinors is trivial [18]
 by virtue of Prop. 5 and its {\sl Corollary}. The
projection $\varpi :\Spin _{m+1}\to \Bbb S _m$
is given by $\varpi (a)=a\vts  e_{m+1}\vts  a^{-1}$, where $(e_1
,\dots  ,e_{m+1})$ is the canonical frame in $\Bbb
R^{m+1}$. Consider a Dirac or a Pauli representation
of $\Spin _m$ in $\GL (S)$ and let $\gamma '$ be
one of its extensions to $\Spin _{m+1}$. For
every  $\varPsi :\Bbb S _m \to S$ the map $\psi
:\Spin _{m+1}\to S$ given by $\psi (a)=\gamma
'(a^{-1})\vts  \varPsi (\varpi (a))$ is a spinor field on the
sphere; every such field can be so obtained. This
observation is implicit in the work of Schr\"odinger
[19] on the Dirac equation on low-dimensional
spheres; see also [20].
\smallskip
\noindent {\sl Remark 1.} Consider the group $\Spin_{m+1}$ as a subgroup 
of $\Pin_{0,m+1}$ or $\Pin_{m+1,0}$.  The map $\varsigma :\Spin_{m+1}\to
\Spin_{m+1}$ given by $\varsigma (a)=
e_{m+1}\vts a\vts e^{-1}_{m+1}$ is an involutive automorphism of 
$\Spin_{m+1}$, preserving the subgroup $\Spin_m$. If $\tau :
\Bbb S_m \to \Bbb S_m$ is the symmetry $x\mapsto e_{m+1}\vts
x\vts e^{-1}_{m+1}$, then $\varpi \circ \varsigma =\tau \circ
\varpi$.  Since $\varsigma (ab)=\varsigma (a)b$ for every $b\in 
\Spin_m$, the pair $(\tau ,\varsigma )$ is an automorphism 
of the spin structure of the sphere and
 if $\psi$ is a spinor field, then so is $\psi_\varsigma =
\psi \circ \varsigma$. If $m$ is even and $b$ is an odd element of
$\Pin_{0,m}$ or $\Pin_{m,0}$, then $\varpi (a\vts 
b\ts\vol_{m+1})=-\varpi (a)$; if $m$ is odd, then 
$\varpi (a\vts\vol_{m+1})=-\varpi (a)$. 
\smallskip
\noindent {\sl Remark 2.} The case of $m$ odd, 
$m=2\nu -1$, deserves an additional 
comment. There are two Weyl representations $\gamma_+$ and 
$\gamma_-$ of $\Spin_{2\nu}$ in $S_+$ and $S_-$, respectively. By 
restriction to  $\Spin_{2\nu -1}$, they give two equivalent Pauli 
representations: there is an isomorphism 
$\theta : S_+ \to S_-$ such that  $\theta \circ \gamma_+ (b) =
\gamma_- (b)\circ \theta$ for every $b\in\Spin_{2\nu -1}$. 
For every $a\in\Spin_{2\nu}$, the linear map $\gamma_- (a)\circ
\theta\circ\gamma_+ (a^{-1})$ is an isomorphism of $S_+$ on 
$S_-$; as a function of $a$ it is constant on the fibers of $\varpi$;
therefore, it defines a map $\Theta :
 \Bbb S_{2\nu -1}\to {\rm Iso}\ts
(S_+ ,S_- )$ such that $\Theta (\varpi (a))\circ \gamma_+ (a)=
\gamma_- (a)\circ\theta$ for every $a\in \Spin_{2\nu}$. 
Since $\varpi (a\vts\vol_{2\nu})=-\varpi (a)$, one has 
$\Theta (-x)=-\Theta (x)$ for every $x\in \Bbb S_{2\nu -1}$. The 
representations $\gamma_\pm$ give rise to the associated bundles 
$\Spin_{2\nu} \times_{\gamma_\pm} S_\pm$ of Pauli spinors ever
$\Bbb S_{2\nu -1}$. According to the {\sl Corollary}, the trivializing 
isomorphisms $\Spin_{2\nu} \times_{\gamma_\pm} S_\pm
\to \Bbb S_{2\nu -1}\times S_\pm$ 
are given by $[(a,\varphi_\pm )]_\pm \mapsto (\varpi (a),
\gamma_\pm (a)\varphi_\pm )$, where $a\in\Spin_{2\nu}$, 
$\varphi_\pm \in S_\pm$ and $[(a,\varphi_\pm )]_\pm =
[(a' ,\varphi'_\pm )]_\pm$ if, and only if,
 there is $b\in\Spin_{2\nu -1}$
such that $a' =ab$ and $\varphi_\pm =
\gamma_\pm (b)\varphi'_\pm$.
The two bundles of Pauli spinors are isomorphic: an isomorphism
$\Spin_{2\nu}\times_{\gamma_+}S_+ \to 
\Spin_{2\nu}\times_{\gamma_-}S_-$ is given 
by $[(a,\varphi_+ )]_+
\mapsto [(a,\theta (\varphi_+ )]_- $ and the 
corresponding isomorphism $\Bbb S_{2\nu -1} \times S_+ \to
\Bbb S_{2\nu -1} \times S_-$ by $(\varpi (a),\varphi_+ )\mapsto
(\varpi (a), \Theta (\varpi (a)\varphi_+ )$.
\bigskip
{\sl B. Real projective spaces.} Recall that the real,
 $m$-dimensional projective space ${\Bbb P }_m$ is
orientable if, and only if, $m$ is odd.  There is the canonical map
$\Bbb S_m \to \Bbb P_m$, $x\mapsto [x]=\{x,-x\}$.
The symmetry $\tau$ of $\Bbb S_m$, defined in {\sl Remark 1},
descends to a symmetry $\tau'$ of $\Bbb P_m$, $\tau' ([x])=
[\tau (x)]$ for $x\in \Bbb S_m$.
 If $k$ is a positive
integer, then $$ {\vol }_{4k}^2 =1,\quad
{\rm but}\quad {\vol }_{4k+2}^2  =-1, $$ and $$
{\vol }_{4k+1,0}^2
={\vol }_{0,4k+3}^2 =1,\quad {\rm but}\quad
{\vol }_{4k+3,0}^2 =
{\vol }_{0,4k+1}^2 =-1. $$ 
To treat simultaneously the spaces $\Bbb P _{4k}$ and
$\Bbb P _{4k+2}$, define
$$
\Pin ^\star
_{2k}=\Pin _{2k,0}\quad{\rm and} \quad 
\vol ^\star_{2k+1}=
\vol _{2k+1,0}\quad {\rm for}\;\; k\;\;{\rm even},
$$
and
$$
\Pin ^\star_{2k}=\Pin _{0,2k}\quad{\rm and}
\quad\vol ^\star_{2k+1}=\vol _{0,
2k+1}\quad {\rm for}\;\; k\;\;{\rm odd}.
$$
For $m$ {\it even\/} there is the monomorphism of
groups $l: \Ort _m \to \SO _{m+1} $
given by $l(A)e_\mu =(\det A)Ae_\mu$ for $\mu
=1,\dots ,m$ and $l(A)e_{m+1}=(\det A)e_{m+1}$. By an
argument similar to the one used in [5] 
to determine the
spin structures on real projective quadrics it follows that:

\ (i) The space ${\Bbb P }_{4k+1}$  has  no spin structure.

(ii) The space ${\Bbb P }_{2k}$ has two inequivalent
$\Pin ^\star_{2k}$-structures $(+)$ and $(-)$, 
$$
\Pin ^\star_{2k}\buildrel{i_\pm}\over{\lto}
\Spin _{2k+1}\to\SO _{2k+1}\to{\Bbb
P}_{2k}, \eqno{(16)}
 $$
  corresponding to the two 
monomorphisms of groups $i_+$ and $i_-$  given by
$$
i_\pm (a)=\cases{
a&for $a\in \Spin _{2k}$,\cr \pm
a\vts\vol ^\star_{2k+1} &for $a =-\alpha (a)
\in \Pin ^\star_{2k}$\cr}$$ 
so that $\rho \circ i_\pm =l\circ \rho$.
The bundle of
Dirac spinors associated with each of the pin structures
on ${\Bbb P }_{2k}$ is trivial: this follows from the
{\sl Corollary\/}  and the observation that the Dirac
representation  of $\Pin ^\star_{2k}$ extends
to the Pauli representation of $\Spin _{2k+1}$.
 The projection $\varpi' :
\Spin_{2k+1}\to \Bbb P_{2k}$ is given by $\varpi'(a)=[\varpi (a)]$.
. The pair $(\tau',\varsigma )$
 is now an isomorphism of one pin structure on 
$\Bbb P_{2k}$ onto the other, as may be seen from 
the easy-to-check equality $\varsigma (a\vts i_+ (b))=
\varsigma (a)\vts i_- (b)$ for every $a\in \Spin_{2k+1}$ and
$b\in \Pin^\star_{2k}$.

(iii) The space ${\Bbb P }_{4k-1}$ has two inequivalent spin
structures, 
$$
 \Spin _{4k-1}\to
\Spin _{4k}/{\bf Z}_2 ^\pm \to
\SO _{4k}/{\Bbb Z}_2 \to {\Bbb P }_{4k-1},
\eqno{(17)}
$$
 where
${\bf Z}_2 ^\pm =\{1,\pm {\vol }_{4k}\}$ and ${\Bbb Z}_2$ 
is the center of $\SO _{4k}$. The
bundle of Pauli spinors associated with each of these
structures is trivial [18]. To see this in detail, let $a\mapsto [a]_\pm 
= \{a, \pm a\vts \vol_{4k}\}$ be the canonical homomorphisms 
of $\Spin _{4k}$ onto $\Spin _{4k}/{\bf Z}_2 ^\pm$. 
 The Pauli representation  of $\Spin _{4k-1}$ 
extends to  representations
$\gamma _\pm '$ of $\Spin _{4k}/{\bf Z}_2 ^\pm$,
 descending from the Weyl representation $\gamma
_\pm$ of $\Spin _{4k}$ in $S_\pm$ such that $\gamma
_\pm({\vol }_{4k})=\pm\id_{S_\pm}$, namely
$\gamma _\pm '([a]_\pm )=\gamma _\pm (a)$. 

The automorphism $\varsigma$ descends to an isomorphism of
groups, $\varsigma':\Spin_{4k}/{\rm \bf Z}^+_2 \to
\Spin_{4k}/{\rm \bf Z}^-_2$, such that $\varsigma'([a]_+ )=
[\varsigma (a)]_-$. The pair $(\tau',\varsigma')$ is 
now an isomorphism of one spin structure on 
$\Bbb P_{4k-1}$ onto the other, but not an equivalence of 
spin structures. 
 The
inequivalence of the two structures described in (ii) and
(iii) is proved in [3].
\eject
\centerline{\bf 5. The Dirac operator}
\bigskip
 \centerline{\it 5.1 Covariant
differentiation of spinor fields}
\bigskip
 Let again (13)
be a pin structure on an $m$-dimensional Riemannian
space $M$. The Levi-Civita connection form on $P$ lifts to
a spin$\vts  (h)$-valued spin connection 1-form $\omega$ on
$Q$.  For every $q\in Q$, there is the orthonormal frame
$\chi (q)=(\chi _\mu (q))\in P$, where $\chi _\mu (q)\in
T_{\varpi (q)}M$ for $\mu =1,\dots ,m$. The spin
connection defines on $Q$ the collection $(\nabla _\mu )$
of $m$ {\it basic} horizontal vector fields such that, for
$\mu =1,\dots  ,m$ and every $q\in Q$, 
$$
\nabla _\mu\vts   \lrcorner\ts  \omega =0\quad{\rm and}\quad
T_q \varpi (\nabla _\mu (q))=\chi _\mu (q).
$$ 
For every $a\in
\Pin (h)$ they transform according to
$$ \nabla _\mu (qa)=T_q \delta (a)\vts  \nabla
_\nu (q)\vts  {\rho ^\nu }_\mu (a). \eqno{(18)}
$$ Let $(e_\mu )$ be a frame in $V$
 and let $\gamma$ be a
spinor representation of $\Pin (h)$ in $S$.
Defining  $\gamma ^\mu =h^{\mu \nu}\gamma _\nu$,
 where $(h^{\mu\nu})$ is the inverse of the matrix $(\langle e_\nu ,h(e_\mu 
)\rangle )$,   and
using Eqs (3) and (4), one obtains
$$ \gamma \circ \alpha (a)\vts  \gamma ^\mu
={\rho ^\mu}_\nu (a^{-1})\vts  \gamma ^\nu\vts   \gamma
(a).\eqno{(19)}
 $$
  Let $\psi: Q\to S$
be a spinor field of type $\gamma$. Its {\it covariant
derivative} is a map $\nabla \psi :Q\to {\Hom}  (V
,S) $ such that, for every  $v=v^\mu e_\mu \in V$, one has
$ \langle v, \nabla \psi \rangle =v^\mu \nabla _\mu \psi $,
where $$ \nabla _\mu \psi =\nabla _\mu \vts  \lrcorner\ts  
\rd \vts   \psi . $$
\bigskip
\centerline{\it 5.2 The classical and the modified  Dirac
operators}
\bigskip
 In the notation of the preceding paragraph, the
{\it classical Dirac operator\/} $D^{\rm cl}$ is given by 
$$
{D^{\rm cl}} \psi =\gamma ^\mu \nabla _\mu \psi . \eqno{(20)}
$$
 According
to (18) and (19), the
classical Dirac operator maps a spinor field of type
$\gamma$ into a spinor field of type $\gamma \circ
\alpha$.

Let $\gamma _{m+1}$ be the isomorphism intertwining the
representations $\gamma$ and $\gamma \circ \alpha$, as
described in Prop. 2. The {\it modified Dirac
operator}, 
$$ D=\gamma _{m+1}{D^{\rm cl}},\eqno{(21)}$$
preserves the type of the spinor field and the
corresponding eigenvalue equation $ D\psi
=\lambda \psi $ is meaningful on non-orientable pin
manifolds [5,9]. To summarize, one has
\proclaim Proposition 6. The classical Dirac operator (20) 
maps a spinor field of type $\gamma$ into a spinor field 
of type $\gamma\circ\alpha$; the modified Dirac operator 
(21) preserves the type of spinor fields.
\par

 If the dimension $m$ of $M$ is {\it
even}, then one can use the vector representation Ad in the
definition of the pin structure on $M$. The classical Dirac
operator preserves then the type of spinor fields and
there is no need for its modification. Using the decomposition of the 
space of Dirac spinors into the sum of spaces of Weyl spinors and the 
form of the Dirac matrices given in part (i) of Prop. 2, one can 
represent the classical and the modified Dirac operators as
 $$ D^{\rm cl} =\left ( \matrix{0
&D^- \cr
 D^+  &0\cr}\right ) \quad {\rm and}\quad  D =\left ( \matrix{0
&\ri D^-\cr
 -\ri D^+ &0\cr}\right ), $$
where $D^\pm$ are the {\it Weyl operators\/}: they act on Weyl spinor 
fields and change their helicity. For an even-dimensional {\it spin\/} 
manifold,  if $\psi$ is a Dirac spinor field, then so is $\gamma_{m+1}\psi$. 
Therefore, the operators $D$ and $-D$ are equivalent, $-D=\gamma_{m+1}
D\gamma^{-1}_{m+1}$, and the spectra of both $D$ and $D^{\rm cl}$ are 
symmetric.  

If $M$ is an {\it odd\/}-dimensional {\it
spin\/} manifold, then the interesting object is the {\it
Pauli operator\/} $D_0 =\sigma ^\mu \nabla _\mu$
 acting on Pauli spinor fields. According to Prop. 2 and 3, one can write
$$D^{\rm cl} =\left (
\matrix{D_0 &0 \cr 0 &-D_0}
 \right )\quad {\rm and}\quad   D =\left (
\matrix{0 &D_0 \cr D_0 &0
} \right ). $$ 
In this case, however, the operators $D_0$ 
and $-D_0$ are not equivalent 
and the spectrum of $D_0$ need
 not be symmetric. Each of the operators $D_0$ and $-D_0$ is `equally 
good'. In other words, the spectrum of an odd-dimensional spin manifold 
is defined only up to `mirror symmetry', $\lambda\mapsto -\lambda$.

Irrespective of the parity of $m$,  if $M$ is
 a spin manifold and $\psi$ is a
Dirac or Cartan
spinor field, then $\gamma_{m+1}\psi$ is a spinor field of
the same type. From (8) one obtains $(1+\gamma
_{m+1})^{-1}=\textstyle {1\over 2} (1- \gamma _{m+1})$ and $ 
D=(1+\gamma _{m+1})\vts   {D^{\rm cl}}\vts  (1+\gamma
_{m+1})^{-1} $ so that if ${D^{\rm cl}} \psi =\lambda \psi$ then
$D \psi '=\lambda \psi '$, where $\psi '=
(1+\gamma _{m+1})\psi$.

\bigskip
\centerline{\bf 6. Pin structures on hypersurfaces}
\bigskip
\centerline{\it 6.1 Existence}
\bigskip

Let $M$ be a hypersurface in  a proper Riemannian spin
$(m+1)$-manifold $N$, defined by an isometric immersion
$f:M\to N $. The hypersurface need not be orientable. The
normal bundle $T^\perp M$ is a line bundle and the Whitney
sum $TM\oplus  T^\perp M$ is isomorphic to the pullback
of $TN$ to $M$ by $Tf$. Since $N$ is spin, its first and
second Stiefel-Whitney classes vanish and the Whitney
theorem  gives $$ w_1 (TM) +w_1 (T^\perp M
)=0\quad{\rm and}\quad w_2 (TM )+ w_1 (TM)w_1
(T^\perp M )=0. $$ Therefore, according to the Karoubi
theorem [1], the hypersurface $M$ has a
$\Pin_{0,m}$-structure. For example, since $\Bbb P _n$ is a
spin manifold for $n\equiv 3\bmod 4$, and the real
projective quadric $Q_{k,l} =(\Bbb S _k \times \Bbb S _l
)/\Bbb Z _2$ is orientable for $k+l$ even, the natural
immersion $Q_{k,l}\to \Bbb P _{k+l+1}$ gives, for
$k+l\equiv 2\bmod 4 $, a spin structure on the quadric with
a proper Riemannian metric, see [4].

According to general theory [8], two immersions of $M$ into $N$, which are 
homotopic one to another,  give rise to 
isomorphic pin structures on $M$.
The circle $\Bbb S_1$ is known to have two inequivalent spin structures: 
the trivial one, $\Spin_1 =\Bbb Z_2\to \Bbb Z_2 \times {\rm U}\vts  _1 \to
{\rm U}\vts  _1 =\SO_2 \to \Bbb S_1$ and the non-trivial structure, 
corresponding to the `squaring' map, $\Spin_2 ={\rm U}\vts  _1 \buildrel{sq}\over{\lto}
{\rm U}\vts  _1 =\SO_2$. Up to homotopy, all immersions of $\Bbb S_1$ into 
$\Bbb R^2$ are classified by integers: with $n\in \Bbb Z,\;n\neq 0$, there is 
associated the class represented by the immersion ${\rm U}\vts  _1\to \Bbb 
C =\Bbb R^2$ given by $z\mapsto z^n$. One can easily verify that the spin 
structures on the circle   corresponding to
immersions with $n$ even (resp., odd) are trivial (resp., non-trivial).

\bigskip
\centerline{\it 6.2 Construction}
\bigskip
 Let $$
\Spin _{m+1}\to Q'\buildrel{\chi
'}\over{\to}P'\buildrel {\pi '}\over{\to}N \eqno{(22)} $$
be a spin structure on $N$ and let $\pi :P\to M$ be the
$\Ort _m$-bundle of all orthonormal frames on
the hypersurface $M$ immersed by $f$ isometrically in
$N$. Define the map $f':P\to P'$ so that if $p=(p_\mu )\in P$
and $x=\pi (p)$, then the frame $f'(p)=(f'_i (p))$ at $f(x)\in
N\vts  \vts   (i=1,\dots ,m+1)$ is given by 
$$
 f'_\mu (p)=(T_x f)(p_\mu )\quad {\rm for}\quad \mu =1,\dots ,m 
$$ 
and
$f'_{m+1}(p)$ is a unit vector at $f(x)$, orthogonal to $T_x
f(T_x M)$ and oriented in such  a way that $f'(p)\in P'$. It is
clear that $f'$ is an injection. Let $\iota
:\Pin _{0,m}\to \Spin _{m+1}$ be as in
(10) for $k=0$ and $l=m$ and let $\kappa
:\Ort _m \to \SO _{m+1}$ be the
corresponding monomorphism of the orthogonal groups,
$\kappa \circ \rho =\rho \circ \iota$. For every $A\in
\Ort _m$ and $p\in P$, one has $f'(pA)=f'(p)\kappa
(A)$. Therefore, the pair $(f,f')$ is a morphism of principal
bundles. The pin$_{0,m}\vts$-structure on $M$ is now given as
a bundle $\chi :Q\to P$ induced from the bundle $\chi '
:Q'\to P'$ by the map $f'$. Explicitly, $$ Q=\{(p,q' )\in
P\times Q' :f'(p)=\chi '(q')\},\quad \chi (p,q')=p. $$ The
action of $ \Pin _{0,m}$ on $Q$ is given by
$$ (p,q')a=(p\rho (a),q'\iota (a)) \eqno{(23)}
$$ so that $\chi ((p,q')a)= \chi (p,q')\rho (a)$
for every $a\in \Pin _{0,m}$.

The pin structure on a hypersurface $M$, constructed in
this manner, is said to be {\it induced\/} by the
immersion $f:M\to N$.

Let $\gamma ' :\Spin _{m+1}\to
{\GL }\vts  (S)$ be a spinor representation and let
$\psi ':Q'\to S$ be a spinor field on $N$ of type $\gamma '$.
It follows from (23) that its restriction $\psi$ to
$Q$, $\psi (p,q')=\psi '(q')$, is a spinor field on $M$ of type
$\gamma =\gamma '\circ \iota$.
\bigskip
\centerline{\it 6.3 Spinors
 on hypersurfaces in Euclidean spaces}
 \bigskip
As an important special case, consider a hypersurface $M$
immersed in the Euclidean space $\Bbb R ^{m+1}$ with its standard flat 
proper-Riemannian metric. 

\proclaim Proposition 7.
 Let $f$ be an
isometric immersion of a hypersurface $M$ in the
Euclidean space $\Bbb R ^{m+1}$.  The pin-structure on $M$,
induced by the immersion, 
$$
 \Pin _{0,m}\to
Q\buildrel{\chi}\over{\to}P\buildrel{\pi}\over{\to} M, 
$$
 is such that
the bundle of Dirac  ($m$ even) or Pauli
($m$ odd) spinors on $M$ is trivial. 
\par

 Since the spin structure (22) of the ambient space reads now
$$ 
\Spin _{m+1}\to \Bbb R ^{m+1}\times
\Spin _{m+1} \to \Bbb R ^{m+1}\times
\SO _{m+1}\to \Bbb R ^{m+1}, 
$$ 
the map $f':P\to
\Bbb R ^{m+1}\times \SO _{m+1}$, defined in the
preceding paragraph, can be written as $f'=(f\circ \pi ,
F)$, where
 $$
  F:P\to \SO _{m+1} \eqno{(24)} 
$$ 
satisfies $F(pA)=F(p)\kappa (A)$
for every $p\in P$ and $A\in \Ort _{m}$. Let
$k:P\to M\times \SO _{m+1}$ be the map $k=(\pi
,F)$. The definition of $Q$ can be simplified to read $$
Q=\{(x,a)\in M\times  \Spin _{m+1}:(x,\rho
(a))\in k(P)\}. $$ Since $k$ is injective, the projection $\chi
:Q\to P$ is well-defined. If $p=\chi (x,a)$, then the
definition of $Q$ implies 
$$
 F(p)=\rho (a). \eqno{(25)} 
$$
 The action of
$\Pin _{0,m}$ on $Q$ is now given by
$(x,a)b=(x,a\iota (b))$, where $(x,a)\in Q$ and $b\in
\Pin _{0,m}$. The bundle $Q\times_\iota
\Spin _{m+1}\to M$, obtained by extending the
structure group $\Pin _{0,m}$ of $Q$ to
$\Spin _{m+1}$, is isomorphic with the trivial
bundle $M\times  \Spin _{m+1}\to M$: an
isomorphism is given by $[((x,a),a')]\mapsto (x,aa')$,
where $(x,a)\in Q$ and $a'\in  \Spin _{m+1}$. The
spinor representation $\gamma ':
\Spin _{m+1}\to {\GL }\vts  (S)$ extends
$\gamma =\gamma '\circ \iota$. For $m$ even (resp., odd),
one takes $\gamma '$ to be a Pauli (resp., Weyl)
representation, so that $\gamma$ is the Dirac (resp.,
Pauli) representation. Applying Prop. 4 to the
present case, with $G= \Pin _{0,m}$ and $G'=
\Spin _{m+1}$, one obtains that the bundle of
spinors $Q\times_\gamma S\to M$ is trivial.
Let $\psi :Q\to S$ be a spinor field of type $\gamma$ on
$M$, i.e. $\psi (x, a\iota (b))=\gamma (b^{-1})\psi (x,a)$
for every $(x,a)\in Q$ and $b\in \Pin _{0,m}$. The
map $Q\to S$, given by $(x,a)\mapsto \gamma ' (a)\psi
(x,a)$, is constant on the fibers of $Q\to M$. There thus
exists a map 
$$
 \varPsi :M\to S \eqno{(26)}
$$ 
such that 
$$ 
\varPsi (x)=\gamma '(a)\psi (x,a) \eqno{(27)} 
$$
 for
every $(x,a)\in Q$. Conversely, for every map (26),
the {\it Schr\"odinger transformation\/} (27)
defines a spinor field $\psi$ of type $\gamma =\gamma
'\circ \iota$ on $M$. An equivalent way of defining the map
$\varPsi$ associated with the spinor field $\psi$ of type
$\gamma$ is to consider the latter's extension $\psi '$ to
the trivial bundle $M\times \Spin _{m+1}\to M$
such that $\psi '(x,a)=\psi (x,a)$ for every $(x,a)\in Q$ and
$\psi '(x,ab)=\gamma '(b^{-1})\psi '(x,a)$ for every $x\in
M$ and $a,b\in \Spin _{m+1}$. If $s:M\to M\times
\Spin _{m+1}$ is the standard section
$s(x)=(x,1)$, then $$ \varPsi =\psi '\circ s.
 $$
\bigskip
\centerline{\bf 7. A formula for the Dirac
  operator on orientable hypersurfaces}
  \bigskip
\centerline{\it 7.1 The general case}
\bigskip
Assume, for simplicity, that the  hypersurface $M$,
immersed isometrically in $\Bbb R ^{m+1}$, is connected,
orientable and has been oriented by distinguishing a
connected component $SP$ of its bundle $P$ of all
orthonormal frames. The pin structure on $M$, induced by
the immersion $f$, can be now restricted to the group
$\Spin _m$ by taking $SQ=\chi ^{-1}(SP)$, see
(13) and (14). 

The map (24)
restricted to $SP$ defines the {\it Gauss map\/}
 $n:M\to \Bbb S _m \subset \Bbb R ^{m+1}$ of
unit normals to $M$. It is given by $ n(\pi
(p))=F(p)e_{m+1} $ and, by virtue of  (25), for
every $(x,a)\in Q$, one has 
$$
 n(x)=a\vts
e_{m+1}\vts a^{-1}, \eqno{(28)}
 $$
  where the product
on the right is given by multiplication in
$\Pin _{0,m+1}$. Let $\gamma$ be a spinor
representation of $\Pin _{0,m+1}$ in $S$; by
restriction, it gives rise to representations of its
subgroups, 
$$
 \Spin _{m} \to
\Spin _{m+1} \to
\Pin _{0,m+1}\buildrel{\gamma}\over{\to}
\GL (S). 
$$ 
Since the first two arrows are
standard injections, there is now no need to  introduce a
separate notation for the restrictions and to distinguish
$\gamma$ and $\gamma '$ as in Prop. 4. In
particular, for every $i=1,\dots ,m+1$, one has the Dirac
matrix $\gamma _i = \gamma (e_i )$ and 
$$
{\rm if}\quad \sigma _{ij} =\gamma _i \gamma _j +\delta
_{ij}, \quad{\rm then}\quad  \sigma _{ij} +\sigma _{ji}=0
\eqno{(29)} 
$$
 for $i,j=1,\dots m+1$.

 The following identities are useful:
$$
\gamma _i \sigma _{jk}=\gamma _{[i}\gamma _j \gamma
_{k]} +\delta _{ik}\gamma _j -\delta _{ij}\gamma _k,
\eqno{(30)} 
$$
$$
\sigma _{ij}\sigma _{kl}=\gamma _
{[i}\gamma _j \gamma _k \gamma _{l]} +\delta _{ik}\sigma
_{jl}-\delta _{jk}\sigma _{il} +\delta _{jl}\sigma _{ik}- 
\delta _{il}\sigma _{jk} +\delta _{il} \delta _{jk}-\delta
_{ik}\delta _{jl},
\eqno{(31)}
$$
where it is understood that there is antisymmetrization
over the
indices included in square brackets.
  By
applying $\gamma$ to both sides of (28), one obtains,
for every $(x,a)\in Q$, $$
\bn(x)= \gamma (a)\gamma _{m+1}\gamma
(a^{-1}),\quad {\rm where}\quad  \bn=\gamma
\circ n ={\textstyle\sum_i}n_i\gamma_i . \eqno{(32)} $$

According to Prop. 6, the modified Dirac operator $ D$ maps a spinor
field $\psi$ of type $\gamma$ into a spinor field of the
same type. There thus exists a linear differential
operator $\cal D$, the {\it Schr\"odinger transform\/} 
of $D$, acting on maps from $M$ to $S$, such that
$$ \gamma (a)(D\psi
)(x,a)=({\cal D}\varPsi )(x), \eqno{(33)} $$ where
$(x,a)\in Q$ and $\varPsi$ is given by (27).
Symbolically, 
$$ {\cal D}=\gamma (a)D \gamma (a^{-1}).
$$ 
Since $D$ anticommutes with $\gamma _{m+1}$,
one obtains from (32) $$
{\cal D}{\bn} +{\bn}{\cal D}=0. \eqno{(34)}
$$
Let $\partial _i$ denote the (constant) vector field on
$\Bbb R ^{m+1}$ given by differentiation with respect to
the $i$th Cartesian coordinate $x_i$, i.e. $\partial _i \varphi =
e_i \vts   \lrcorner\ts   \rd \varphi$ for every function
$\varphi$ on  $\Bbb R ^{m+1}$. The field of unit normals,
$n= \sum_i n_i \vts  \partial _i$, defines a collection of
$\textstyle {1\over 2} m(m+1)$ vector
 fields $n_i \partial _j -n_j
\partial _i\;\; (1\leq i<j\leq m+1)$ tangent to the
hypersurface.

To determine  explicitly the Schr\"odinger transform $\cal D$ of 
the modified Dirac
operator, consider the extension $\psi '$ of a spinor 
field of type $\gamma$ on $M$ defined at the end of Sec. 
6.3.  The connection form on $SQ$ 
extends to a ${\rm spin}_{m+1}\vts$-valued connection form 
$(\omega '_{ij})$ on $M\times \Spin_{m+1}$. With the 
conventions of Eq. (29), the covariant exterior derivative 
of $\psi '$ is
 $$
{\rm hor\ d}\psi ' = \rd \psi ' -{\textstyle 
{1\over 4} \sum} _{i,j}\sigma _{ij} \vts  \omega ' _{ij}\vts  \psi ' .
$$
The connection form  pulled back to $M$ by the standard 
section $s$ is 
$$
\omega_{ij} =s^* \omega '_{ij} =n_i \vts  {\rd}n_j -n_j \vts  {\rd }n_i .
$$
Let $\varepsilon =\rd x_1 \wedge \dots \wedge \rd x_{m+1}$ 
denote the canonical volume form on $\Bbb R^{m+1}$ and let 
$\varepsilon _i =e_i \vts  \lrcorner \ts  \varepsilon$ so that 
$\rd \psi \wedge \varepsilon _i =\varepsilon \vts  \partial _i 
\psi$ and $\omega _{kl} \wedge \varepsilon _i =(n_k 
\partial _i n_l -n_l 
\partial _i n_k )\vts  \varepsilon$ on $M$. Noting that the $m$-form 
$n\vts\lrcorner\ts\varepsilon$ is the volume 
form of the hypersurface, using an expression 
of the Dirac operator with the help of differential forms 
[9,24] and pulling it back to $M$ by the section $s$, one 
obtains 
$$
(n\vts  \lrcorner\ts  \varepsilon ){\cal D}\varPsi ={\textstyle 
\sum} _{i,j}\sigma_{ij}\vts  n_i\vts  n\vts  \lrcorner\ts  
((\rd \varPsi 
-{\textstyle{1\over 4}}{\textstyle\sum} _{k,l}\sigma _{kl}\vts 
 \omega _{kl}\vts  \varPsi )
\wedge \varepsilon _j ).
$$
With the help of  Eqs (30) and (31) one finds
$$
{\cal D} ={\textstyle\sum} _{i,j} 
\sigma _{ij}\vts  n_i \vts  \partial _j +\textstyle {1\over 2}
{\rm div}\vts  n, \eqno{(35)}
 $$
  where the
`intrinsic divergence' $ {\rm div}$ is given by 
$$
{\rm div}\vts  n={\textstyle\sum} _{i,j}(\delta _{ij}-n_i 
\vts  n_j)\vts  \partial _i\vts  n_j . 
$$

The differential operator $\sum_{i,j} \sigma _{ij}n_i\vts  
\partial _j$ has been studied by Delanghe and Sommen
[21] who refer it to an
unpublished thesis by Lounesto; see also [22,23] and
the bibliography given there.  The divergence term in (35) is essential 
for $\cal D$ to correspond to the intrinsic (modified) Dirac operator on $M$.
In the special case when $M$ is the hyperplane given by $x_{m+1}=0$, one 
has $n_i =\delta_{i,m+1}$ and $\cal D$ 
reduces to $\gamma_{m+1}\sum_{\mu =1}^m \gamma_\mu \partial_\mu$.
\bigskip
\centerline{\it 7.2 The case of odd-dimensional hypersurfaces}
\bigskip
If the hypersurface $M$ is odd-dimensional, $m=2\nu -1$, 
and orientable, then it is enough to consider its bundle 
of Pauli spinors, which is of fiber dimension $2^{\nu -1}$. Let 
 $\gamma_i 
,\;i= 1,\dots ,2\nu$,  be the Dirac matrices associated 
with the ambient space $\Bbb R^{2\nu}$ and put 
$\gamma_{2\nu +1}= \gamma_1 \dots \gamma_{2\nu}$. The matrix
$\gamma_{2\nu +1}$ is unchanged by the Schr\"odinger 
transformation; its eigenvectors are Pauli spinors and  
 the space of Cartan 
spinors $S$, associated with $M$, decomposes into the sum 
$S_+ \oplus S_-$ of spaces of Pauli spinors. Since $\gamma_{2\nu +1}$ 
commutes with the operator $\cal D$
 (and also with $D$), it suffices to consider eigenfunctions 
of $\cal D$ with values in one or the other space of Pauli spinors; see 
  {\sl Remark 2\/} in Sec. 4.3; it applies, {\it mutatis mutandis}, to all
odd-dimensional hypersurfaces in $\Bbb R^{m+1}$.
Taking the Dirac matrices $\gamma_i$ in the form described 
in part (i) of Prop. 2 one can write
$$ \bn =\left ( \matrix{0
&\bn^-\cr
\bn^+&0\cr}\right ), $$
where, for every $x\in \Bbb R^{m+1},
\; \bn ^\pm (x): S_\pm \to S_\mp$. 
The matrices $\sigma_{ij}$, corresponding to even elements of the 
Clifford algebra, preserve the helicity,
$$ \sigma_{ij} =\left ( \matrix{\sigma^+_{ij}
&0\cr
0&\sigma^-_{ij}\cr}\right ). $$  The same is true of the Dirac operator,
$$ {\cal D} =\left ( \matrix{{\cal D}^+
&0\cr
 0&{\cal D}^- \cr}\right ),$$ 
and (34) gives 
$$ {\cal D}^\pm \bn ^\mp +\bn ^\mp {\cal D}^\mp =0.
$$
If $\varPhi :M\to S_+$ is an eigenfunction 
of ${\cal D}^+$ with eigenvalue 
$\lambda $, then $\bn ^+ \varPhi :M\to S_-$
 is an eigenfunction of ${\cal 
D}^-$ with eigenvalue $-\lambda$.
\bigskip
\centerline{\it 7.3 The case of a foliation 
of $\Bbb R^{m+1}$ by hypersurfaces}
\bigskip
  Consider  an open
subset $U$ of $\Bbb R ^{m+1}$ foliated by a family of
hypersurfaces. The Gauss map defines now a field $n$ on
$U$. Let $\partial /\partial r =\sum_i n_i \partial _i$ be
the derivative along $n$. Define the classical Dirac
operator in $\Bbb R ^{m+1}$ as
$$\bpartial={\textstyle\sum}_i \gamma _i \vts  \partial _i .$$
 A
simple computation, based on (29) gives
$$ \bn\vts  
{\bpartial}={\cal D} -( {{\partial}\over {\partial r}}
+{1\over 2} {\rm div}\vts   n). \eqno{(36)}
$$
Similar formulae, expressing the split of the Dirac operator $D$ on a 
manifold with boundary into parts tangential and transversal to the 
boundary, are used in the index theory of $D$ [7,25].

For $m$ odd, in the notation of Sec. 7.2, one has
$$ \bpartial =\left ( \matrix{0
&\bpartial^-\cr
\bpartial^+&0\cr}\right ). $$
Eq. (36) gives now
$$
\bn ^\mp \bpartial ^\pm ={\cal D}^\pm  -({\partial 
\over {\partial r}}+ {1 \over 2}{\rm div}\,n) .
$$
\eject
\centerline{\bf 8 Applications}
\medskip
\centerline{\it 8.1 The spectrum of the sphere $\Bbb S_m$}
\medskip
 Let the integer $m$
be $>1$.
The set $U=\{x=(x_i )
\in \Bbb R ^{m+1}: x\neq 0\}$ is foliated by the spheres
$r={\rm const.}>0$, where $r=(x_1 ^2 +\dots
+x_{m+1}^2 )^{1/2}$. Since now $n_i =x_i /r$, one has
${\rm div}\vts  n=m/r$. The differential operators 
$r(n_i \partial _j -n_j \partial _i )$ generalize the
operator $\vec{L}=\vec{r}\times \vec{p}$ of `orbital
angular momentum' and  the operator
$r\sum_{i,j} \sigma _{ij}n_i \partial _j$ corresponds to the
`spin-orbit coupling' term $\vec{\sigma}\vec{L}$
of
quantum mechanics in $\Bbb R ^3$. 
If $r=1$, then,  by virtue of (31),
 $$ ({\textstyle\sum_{i,j}}
\sigma _{ij}x_i \partial _j )^2 +(m-1)
{\textstyle  \sum_{i,j}} \sigma _{ij}x_i \partial _j +\Delta =0,
$$
where $\Delta$ is the Laplace operator on $\Bbb S _m$.
Together with (35), this gives 
$$ ({\cal D} -\textstyle {1\over 2} m)({\cal D} +
\textstyle {1\over 2} m-1)+\Delta =0.
$$
If $\varPsi :\Bbb S _m \to S$ is an eigenfunction
 of ${\cal D}$, then
it is also an eigenfunction of $\Delta$. Every eigenvalue of
$\Delta$ on $\Bbb S _m$ is known to be of the form
$-l(l+m-1)$ for some $l=0,1,\dots$. Therefore, if
${\cal D} \varPsi =\lambda \varPsi$, then $(\lambda
-\textstyle {1\over 2} m)(\lambda +
\textstyle {1\over 2} m -1)=l(l+m-1)
$, i.e. either $\lambda =l+\textstyle {1\over 2} m$
 or $\lambda 
 =-l-\textstyle {1\over 2} m +1$. If ${\cal D} \varPsi
 =(-\textstyle {1\over 2} m+1)\varPsi
 $, then $\Delta \varPsi =0$; therefore, $\varPsi$ is
 a constant and (35) gives ${\cal D} \varPsi
 =\textstyle {1\over 2} m\varPsi$. Since $m\neq 1$,
  the equality 
 $(-\textstyle {1\over 2} m+1)\varPsi =
 \textstyle {1\over 2} m\varPsi$ implies
 $\varPsi =0$ and so
 the  number  $-\textstyle {1\over 2} m  +1$ 
  is  not  an eigenvalue:
   the spectrum of ${\cal D}$ on $\Bbb
 S_m$ is  contained  in the set $\{\pm
 (l+\textstyle {1\over 2} m):l=0,1,\dots\}$.
\medskip
\noindent (i) $m$ {\sl even.}
 To show that every element of this set is an eigenvalue
 and to compute its multiplicity, assume first that $m$ is {\it even}, 
$m=2\nu$ and let $S$ be the $2^\nu$-dimensional space of 
spinors. Consider the space
 $H_{m,l}(S)$ of $S$-valued harmonic polynomials on $\Bbb
 R^{m+1}
 $, homogeneous of degree $l$.
 Since $H_{m,l}(S)=H_{m,l}(\Bbb C )\otimes S$, one has
 $$
  \dim H_{m,l}(S) =(s_{m,l}
 +s_{m,l+1})\dim  S,
\eqno{(37)}
 $$
   where
 $$s_{m,l} = \textstyle{{m+l-1}\choose{l}}.$$
 
 Let 
 $\bx$ 
 be the linear map of multiplication of a function $\varPhi :\Bbb 
 R^{m+1}\to S$ by $\sum_i \gamma _i
 x_i $, i.e. $(\bx\vts  \varPhi )(x) =\sum_i \gamma _i
 x_i \vts  \varPhi (x)$. The validity of the following identity is easy to check:
$$
\bpartial \vts \bx ^2 -\bx ^2 \vts \bpartial =2\bx . \eqno{(38)}
$$
If $\varPhi$ is a harmonic polynomial of degree 
 $l$, then the functions $\bpartial(\bx\vts  \varPhi )$ 
 and $\bx\vts  (\bpartial\vts  \varPhi )$ are also  such polynomials. 
\proclaim Lemma 1. For every $\varPhi \in H_{m,l} (S)$ one has
 $$ 
 ({\bpartial}\vts
 \bx+\bx\vts 
 {\bpartial})\vts \varPhi =-(2l+m+1)\vts \varPhi ,\eqno{(39)}
 $$
$$\bpartial (\bpartial \vts\bx  +2)\vts\varPhi =0.\eqno{(40)}
$$
\par
\noindent {\sl Proof.} Since $\varPhi$ is 
homogeneous of degree $l$, the 
Euler identity reads  $\sum_i x_i \partial_i \varPhi =l\varPhi$. 
Using Eq. (29), one obtains 
$${\textstyle\sum}_{i,j}\gamma_i \,\gamma_j (\partial_i x_j +x_i 
\partial_j )\varPhi 
={\textstyle\sum}_{i,j} 
(\sigma_{ij}-\delta_{ij})(\delta_{ij}+x_j \partial_i +x_i \partial _j 
)\varPhi =-(2l+m+1)\varPhi .
$$
Since $\varPhi $ is harmonic, $\bpartial ^2 \varPhi =0$, and
$$
\bpartial ^2 (\bx\ts\varPhi 
)=-{\textstyle\sum}_{i,j}\partial_i ^2 \ts (x_j 
\gamma_j \varPhi )=
-2{\textstyle\sum}_i \gamma_i \ts\partial _i \ts\varPhi 
=-2\bpartial \ts\varPhi .\eqno{ \square}
$$
\smallskip
\proclaim Lemma 2. 
The sequence
$$
\dots \buildrel{\bpartial}\over
{\to}H_{m,l+1}(S) \buildrel{\bpartial}\over
{\to}H_{m,l}(S) \buildrel{\bpartial}\over
{\to}H_{m,l-1}(S) \buildrel{\bpartial}\over{\to}
\dots
\buildrel{\bpartial}\over{\to}S
\buildrel{\bpartial}\over{\to}0
\eqno{(41)}
$$
is exact and there is a decomposition 
$$
 H_{m,l}=H'_{m,l}(S)\oplus H''_{m,l}(S),\eqno{(42)}
$$
where
$$H'_{m,l}(S)=\{\varPhi \in H_{m,l}(S): {\bpartial}
\varPhi =0\}
$$
is the kernel of $\bpartial$ and
$$
H''_{m,l}(S)=\{ \bx\varPhi :\varPhi \in
H'_{m,l-1}(S)\}.
$$
\par
\noindent {\sl Proof.} To show that the sequence is exact, one notices 
that $\bpartial\vts H_{m,l+1}(S)\subset H'_{m,l}(S)$; if $\varPhi\in 
H'_{m,l}(S)$, then, by (39), $\varPhi =-(2l+m+1)^{-1}\bpartial (\bx 
\varPhi )$, i.e. $H'_{m,l}(S)\subset \bpartial\vts H_{m,l+1}(S)$.
 By virtue of Eq. (40), the vector 
space $H''_{m,l} (S) $ is a subspace of $H_{m,l} (S)$ and
 the map $\bx :H'_{m,l-1}(S)\to H''_{m,l}(S)$ 
is an isomorphism of vector spaces.
The sum (42) is direct because   
if $\varPhi \in H'_{m,l}(S)\cap H''_{m,l}(S)$, 
then Eqs (38) and (39) give $2\varPhi =-(2l+m+1)\varPhi $, thus 
$\varPhi =0$.  To show that 
 $H_{m,l}\subset H'_{m,l}(S)\oplus H''_{m,l}(S)$
one writes, as a consequence of (39),
$$
(\bpartial\vts\bx +2)\vts\varPhi +\bx\vts\bpartial 
\vts\varPhi =-(2l+m-1)\varPhi .
$$
According to (40), $(\bpartial\vts\bx +2)\vts\varPhi\in 
H'_{m,l}(S)$. Since $\varPhi$ is harmonic, $\bx\vts\bpartial 
\vts\varPhi $ is in $H''_{m,l}(S)$. Moreover, $m>1$ and 
	$l\geq 0$ imply $2l+m-1>0$. {\hfill $\square$}
\smallskip
By
virtue of (36), if $\varPhi ' \in H'_{m,l}(S)$,
then the
restriction $\varPsi '$ of 
 $\varPhi '$
to the unit sphere  is an eigenfunction of ${\cal D}$,
 $${\cal D}\varPsi '
=(l+\textstyle {1\over 2} m)\varPsi ' \quad{\rm and}\quad \varPsi '
(-x)=(-1)^{l} \varPsi ' (x). $$ for $l=0,1,\dots$.

Similarly, if $\varPhi '' \in H''_{m,l+1}(S)$ then, by virtue
of (39), 
  the restriction $\varPsi '' $ of $\varPhi ''$ 
 to the
unit sphere satisfies
 $${\cal D}\varPsi '' =-(l+\textstyle {1\over 2} m)\varPsi ''
\quad{\rm and}\quad \varPsi '' (-x)=(-1)^{l+1}
 \varPsi '' (x) $$
for $l=0,1,\dots$. 

According to Lemma 2 the vector spaces $H'_{m,l}(S)$ and 
$H''_{m,l+1}(S)$ are isomorphic and
$$
\dim H_{m,l+1}(S) =\dim H'_{m,l}(S) +\dim H'
_{m,l+1}(S). \eqno{(43)}
$$

Writing $\dim H'_{m,l}(S)=p_{m,l} \dim S$
 and using
(37) and (43), one obtains
$$
p_{m,l}+p_{m,l+1}=s_{m,l}+s_{m,l+1}.
$$ 
Since $p_{m,0}=s_{m,0}=1$, one has $p_{m,l}=s_{m,l}$ for
every $l=0,1,\dots$.

\medskip
\noindent (ii) $m$ {\sl odd.}
If $m=2\nu +1$, then $S=S_+ \oplus S_-$ and the spaces of 
Pauli spinors $S_\pm$ are $2^\nu$-dimensional. Spinor fields on $\Bbb 
S_{2\nu -1}$ can be identified with maps from $\Bbb 
S_{2\nu -1}$ to one of the spaces of Pauli spinors, say $S=S_+$, but it 
is convenient to 
consider   sequences such as
$$
\dots 
{\to}H_{m,l+1}(S_- ) \buildrel{\bpartial^- }\over
{\lto}H_{m,l}(S_+ ) \buildrel{\bpartial^+}\over
{\lto}H_{m,l-1}(S_- ) \to
\dots
$$
They are used to  prove suitable modifications of Lemmas 1 and 2. For 
example, $H'_{m,l}(S_+ )$ is now defined as the kernel of 
$\bpartial^+$ and $H''_{m,l}(S_+ )$ as the image of 
$H'_{m,l-1}(S_- )$ by $\bx^-$.
\medskip

Irrespective of the parity of $m$, every
eigenfunction of $\Delta$ on  $\Bbb S _m$ 
is known to be the restriction of a harmonic 
polynomial in $\Bbb R ^{m+1}$; therefore, every
eigenfunction of ${\cal D}$ on $\Bbb S _m$ belongs 
to the restriction of   either
$H'_{m,l}(S)$ or $H''_{m,l}(S)$ for some $l=0,1,\dots$. 
To summarize, one has

\proclaim Proposition 8.
Let $m=2\nu $ (resp.,   $m=2\nu +1$), where $\nu$ is a 
positive integer.
 The spectrum  of the Dirac (resp., Pauli)
 operator  on  $\Bbb S _m$
 is the set
$ \{\pm
(l+\textstyle {1\over 2} m):l=0,1,\dots \}.
$
Each of the eigenvalues $l+\textstyle {1\over 2} m$ and $-l-
\textstyle {1\over 2} m$ occurs with the multiplicity 
$2^\nu {{m+l-1}\choose{l}}$. Let $S$ be the $2^\nu$-dimensional
space of Dirac (resp., Pauli) spinors and
let $\varPsi :\Bbb S _m \to S$ be an eigenfunction of the 
Schr\"odinger transform of the modified
Dirac operator. If
 ${\cal D}\varPsi = ( l+\textstyle {1\over 2} m)\varPsi$, then 
 $\varPsi $ is the restriction of ${\partial}
 (\bx\varPhi )$
to $\Bbb S _m$, where $\varPhi$ is an $S$-valued harmonic
polynomial on $\Bbb R ^{m+1} $, homogeneous of degree $l$.  
If   ${\cal D}\varPsi =- (
l+\textstyle {1\over 2} m)\varPsi$, then
 $\varPsi $ is the restriction
 of $\bx (
 {\bpartial}\varPhi )$, where $\varPhi$ is a
 similar  polynomial    of  degree  $l+1$. 

The same result on the spectrum and its multiplicity is quoted in [26] 
and derived in [27], by a method different from 
the one presented here; see also [28]. 
\bigskip

\centerline{\it 8.2 Application to real projective spaces}
\bigskip

The simple description of the spectrum of the Dirac
operator on spheres, given in the preceding paragraph,
can be used to find the corresponding results for real
projective spaces. Since a real projective space is locally isometric to 
its covering sphere, a spinor field on $\Bbb P_m$ is an eigenfunction of 
the  Dirac operator $D$ if, and only if, it descends from a corresponding 
eigenfunction on $\Bbb S_m$. Therefore, the spectrum of $D$ on $\Bbb P_m$ 
is contained in that of $D$ on $\Bbb S_m$.
\smallskip

 By
comparing (15) with (16),
 one sees that the total
space $\Spin _{2\nu +1}$ defining the pin
structures on $\Bbb P _{2\nu }$ is the same as the total space
defining the spin structure on $\Bbb S _{2\nu } $. Let $\gamma
:\Spin _{2\nu +1}\to {\GL }\vts  (S)
$ be the Pauli representation and let 
$\varpi' :\Spin _{2\nu +1} \to\Bbb P _{2\nu }$ be the
projection defined in part B (ii) of Sec. 4.3.
If $\varPhi :\Bbb P _{2\nu }\to S$, then 
$\psi :\Spin _{2\nu +1}\to S$ defined by $\psi (a)=
\gamma (a^{-1})\varPhi (\varpi' (a))$ is a spinor field of
type $\gamma \circ i_\pm$. Referring to the last sentence of {\sl Remark 
1}, one sees that every spinor field on
$\Bbb P _{2\nu }$ can be so obtained. Therefore, every spinor
field on $\Bbb P _{2\nu }$ comes from an {\it even\/}
function $\varPsi :\Bbb S _{2\nu }\to S$. A similar analysis applies to the 
case $m\equiv 3\bmod 4$. 
 
\proclaim Proposition 9.
Let $m=2\nu$, where $\nu$ is a positive integer (resp., $m=2\nu +1$, 
where $\nu$ is a positive odd integer). The spectrum of the Dirac (resp., 
Pauli)  operator on $\Bbb P _m$
is the set 
$$
\{-\textstyle {1\over 2} \pm (2l+
\textstyle {1\over 2}m+\textstyle {1\over 2}):l=0,1,\dots\}.
$$
The eigenvalue $\lambda$ occurs in the spectrum with the
multiplicity 
$$2^\nu {{|\lambda |+\textstyle {1\over 2}m -1}
\choose{|\lambda |-\textstyle {1\over 2}m}}.$$
\par
Note that, in this case, the spectrum is asymmetric.
\bigskip

  \centerline{\bf Acknowledgments}
  \bigskip

   Most of the work reported
   in this paper was done during my numerous visits to Trieste
and Brussels. I thank Paolo Budinich, Pawe{\l} Nurowski,
Michel Cahen, and Simone Gutt for their interest and 
helpful discussions.

\bigskip
\centerline{\bf References}
\bigskip

\frenchspacing

\item{\ [1]} M. Karoubi,  {\it Ann. Sci. \'Ec. Norm. Sup. \/} 
{\bf 1}, 161 (1968).

\item{\ [2]} S. Carlip, C. DeWitt-Morette,
 {\it
Phys. Rev. Lett.\/} {\bf 60}, 1599  (1988).

\item{\ [3]} L. D\c abrowski, A. Trautman,
 {\it
J. Math. Phys.\/} {\bf 27}, 2022 (1986).

\item{\ [4]} M. Cahen, S. Gutt, A.
Trautman,  {\it J. Geom.
Phys.\/}  {\bf 10}, 127 (1993).

\item{\ [5]} M. Cahen, S. Gutt, A.
Trautman, paper submitted to  {\it J. Geom.
Phys.\/} (1995).

\item{\ [6]} J. Rzewuski, {\it Field Theory}, Part I and II,
PWN, Warszawa 1958 and 1969.

\item{\ [7]} M. F. Atiyah, R. Bott, V. K. Patodi,
 {\it Invent.
Math.\/} {\bf 19}, 278 (1973).

\item{\ [8]} H. B. Lawson, Jr., M. L. Michelsohn,
 {\it Spin Geometry}, Princeton University Press, Princeton
1989.

 \item{\ [9]} A. Trautman, {\it J. Math.
Phys.\/} {\bf 33}, 4011 (1992).
 
 \item{[10]} P. Budinich, A. Trautman,  {\it The
spinorial chessboard}, Trieste Notes in Physics,
Springer--Verlag, Berlin 1988.

\item{[11]} H. Baum, {\it Spin-Strukturen und
Dirac-Operatoren \"uber pseudoriemannschen
Mannigfaltigkeiten}, Teubner-Texte zur Mathematik, 
Teubner, Leipzig 1981.

\item{[12]} \'E. Cartan, {\it Le\c cons sur la th\'eorie
des spineurs}, Hermann, Paris 1938.

\item{[13]} R. Brauer, H. Weyl, {\it Amer. J. Math.\/}
 {\bf 57}, 425 (1935).

\item{[14]} C. Chevalley, {\it The algebraic theory
of spinors}, Columbia University Press, New York 1954.

\item{[15]} L. D\c abrowski, {\it Group Actions on
Spinors}, Bibliopolis, Naples 1988.

\item{[16]} W. Pauli,
{\it Ann. Inst. Henri Poincar\'e\/} {\bf6}, 109 (1936).

\item{[17]} M. F. Atiyah, R. Bott, A. Shapiro,
 {\it Topology\/} {\bf 3} Suppl. {\bf 1}, 3 (1964).

\item{[18]} S. Gutt,  in: {\it Spinors in
Physics and Geometry}, Eds A.
Trautman and G. Furlan, World Scientific, Singapore 1988, 
p. 238.
 
\item{[19]} E. Schr\"odinger, {\it Acta Pontif. Acad. Sci.\/}
 {\bf 2}, 321 (1937).

\item{[20]} W. Pauli, {\it Helv. Phys. Acta\/} {\bf 12},
147 (1939).

\item{[21]} R. Delanghe, F. Sommen, in: {\it Clifford
Algebras and Their Applications in Mathematical Physics},
Eds J. S. R. Chisholm and A. K. Common, Reidel, Dordrecht
1986, p. 115.

\item{[22]} R. Delanghe,  F. Sommen, V. Sou\v
cek, {\it Clifford Algebras and Spinor-Valued Functions},
Kluwer, Dordrecht 1992.

\item{[23]} J. Cnops, {\it Hurwitz pairs and applications of M\"obius 
transformations}, Habilitation Thesis, Ghent University, Ghent 1994.  

\item{[24]} A. Trautman, {\it Symp. Math.\/} {\bf 12}, 139 (1973).

\item{[25]} B. Boo\ss -Bavnbek, K. Wojciechowski, {\it Elliptic Boundary 
Problems for  Dirac Operators}, Birkh\"auser, Basel 1993.

\item{[26]}  P. Van Nieuwenhuizen, 
in: {\it Relativity, Groups and Topology II\/}, 
Eds B. S. DeWitt and R. Stora, North-Holland, Amsterdam 1984, p. 823.

\item{[27]} S. Sulanke, {\it Berechnung des Spektrums des Quadrates des 
Dirac-Operators auf der Sph\"are}, Ph. D. thesis, 
Humboldt-Universit\"at, Berlin 1978.

\item{[28]} A. Trautman,  in: {\it Spinors, Twistors, Clifford
Algebras and Quantum Deformations}, 
 Eds Z.  Oziewicz {\it et al.}, Kluwer,
Dordrecht 1993, p. 25.

\end